\documentclass[12pt]{article}

\usepackage{amsmath}
\usepackage{amssymb}
\usepackage{axodraw}

\begin{document}
\providecommand{\ads}{\text{AdS}}
\providecommand{\btil}{{\tilde b}}
\providecommand{\ftil}{{\tilde f}}
\providecommand{\Ztwo}{\mathbf{Z}_2}
\providecommand{\Gam}{\Gamma}

\begin{flushright}
HU-EP-00/21 \\
\today
\end{flushright}

\vspace{1cm}
\begin{center}
\baselineskip25pt
{\Large\bf Critical Vacuum Energy, Warped Geometry and Grand Unification}
\end{center}
\vspace{1cm}
\begin{center}
\baselineskip12pt
{Axel Krause\footnote{e-mail: {\tt akrause@fas.harvard.edu},
now at Harvard University, Physics Department, Cambridge, MA
02138, USA}}
\vspace{1cm}

{\it Humboldt-Universit\"{a}t zu Berlin,\\[2mm]
Institut f\"{u}r Physik, D-10115 Berlin, Germany}

\vspace{0.3cm}
\end{center}
\vspace*{\fill}

\begin{abstract}
We explore a mechanism to obtain the observational small value for
the 4-dimensional vacuum energy through an exponential warp-factor
suppression. Intriguingly the required suppression scale relates
directly to the GUT scale. We demonstrate the mechanism explicitly
in a 5-dimensional brane-world setup with warped geometry. Upon
lifting the setup to 10-dimensional IIB string-theory, the
relevance of the GUT scale becomes clear as the IIB string-theory
description, which is based on D3-brane stacks, gives rise to a
spontaneously broken SU(5) supersymmetric GUT theory with
low-energy MSSM spectrum and Higgs doublet-triplet splitting.
\end{abstract}

\noindent
\vspace*{\fill}


\section{Introduction}

The enormous smallness of the 4-dimensional vacuum energy,
constrained by cosmological and astronomical measurements to be
\cite{Bahcall}
\begin{equation}
   |\Lambda_4| \lesssim 10^{-47} \text{GeV}^4
               = (1.8\,\text{meV})^4 \; ,
     \label{Exp}
\end{equation}
is still not understood in a satisfactory way from a theoretical
point of view. The energy-regime of the upper bound of some meV is
rather unnatural in particle physics and more characteristic of
condensed matter phenomena. However, it has to be noticed that the
upper bound on the electron neutrino mass can be as low as 1meV
\cite{Klapdor}, which comes strikingly close to this value. If
experiment will eventually show that both numbers are indeed so
close, it would be an intriguing hint to some deeper relation
between the Standard Model (SM) and gravity.

The hope that eventually a consistent theory of quantum gravity
might be able to explain the vexing smallness of the vacuum energy
resp.~cosmological constant has not been fulfilled yet, as the
leading candidate, M/string-theory, relies so heavily on exact
supersymmetry. Since the tininess of the cosmological constant is
measured at energies where Bose-Fermi degeneracy is seen to be
violated, a supersymmetry-breaking mechanism would be needed which
nonetheless should not give rise to a large $\Lambda_4$. An
interesting M-theory inspired proposal has been made in
\cite{WittenCosm}. The idea is that in three dimensions,
supersymmetry enforces a zero cosmological constant but can exist
without matching bosonic and fermionic degrees of freedom. If such
a 3-dimensional theory contains a modulus similar to the dilaton
of string-theory, one could expect that at strong coupling an
additional fourth dimension will open up much like in M-theory.
The hope would be that during the transition from weak to strong
coupling the zero cosmological constant and Bose-Fermi
non-degeneracy might be preserved. Another interesting aspect
which arose in string-theory is that vacua with zero and negative
cosmological constant can sometimes be connected via T-duality
\cite{CCT}. This suggests that vacua with negative cosmological
constant might in fact be viewed as flat spacetime vacua. Again
this connection has so far only been found in three dimensions.
Finally, there might also be a radically different understanding
of the vacuum energy if M-theory turns out to be a theory of only
a finite \cite{Banks}, \cite{KCT} but huge amount of discrete
degrees of freedom as suggested for instance by microscopic
entropy considerations.

Whereas in the very early universe a large positive cosmological
constant is welcome during the phase of inflation, we face the
problem to understand the smallness of the cosmological constant
in our low-energy world today. Therefore, we shall take the point
of view in this paper, that there should also exist a rationale to
understand the adjustment of the cosmological constant to tiny
values not only by taking refuge to a quantum gravity description
valid at Planck-energies but also by employing merely degrees of
freedom which are available at low energies.

Furthermore, we shall adopt the view that our 4-dimensional world
arises as a brane-world from stacks of branes, embedded in some
higher-dimensional spacetime. Conceiving our world as being
located on a type IIB string-theory D3-brane in a 10-dimensional
ambient space allowed to attack such fundamental problems as gauge
and gravitational coupling unification or the Standard Model
hierarchy problem from completely different point of views (see
\cite{KakushadzeTye} and references therein) than the traditional
technicolor or low-energy supersymmetry approaches. In a
T-dualized type I string scenario, where two to six internal
compact dimensions orthogonal to the D-brane are chosen much
larger than the remaining compact dimensions, one is able to lower
the fundamental higher-dimensional Planck scale down to the TeV
scale \cite{Antoniadis}. This necessitates the large internal
dimensions to be as large as 1mm resp.~1 fermi for two resp.~six
large internal dimensions. Most pronounced in the case of two
large dimensions, this leads to another hierarchy between the new
fundamental TeV scale and the compactification scale $\mu \equiv
\hbar c/1\text{mm} \approx 10^{-4}$eV. This drawback could be
overcome by considering not a direct product structure for the
background space-time but a warped metric instead. In particular,
the warped metric of a slice of an AdS-space suspended between two
branes offers a solution to the strong part of the hierarchy
problem \cite{RS1}.

In \cite{GoldWise} it has been shown how to stabilize the modulus,
which describes the distance between the two branes, at a value of
10-50 Planck lengths. This is the value which is compatible with
the mentioned solution of the hierarchy problem. It remains to
relax the fine-tuning condition between the bulk cosmological
constant and the brane-tensions. Attempts in this direction have
been undertaken recently (see e.g.~\cite{ADKS}, \cite{KSS},
\cite{FLLN1}). However, the solution to the hierarchy problem
cannot be maintained in these approaches as the solutions exhibit
metrics that show polynomial instead of exponential behaviour. The
metrics vanish at two finite points in the extra dimension,
thereby cutting off the infinite range through singularities.
However, the resolution of these singularities remains obscure.

A general review of the cosmological constant problem can be found
in \cite{Weinberg1}. See \cite{Weinberg2}, \cite{Carroll} for more
recent reviews on the topic. \cite{WittenRev} provides a recent
discussion of the cosmological constant problem from the point of
view of string-theory.  Apparently, lately there has been a
noticeable increase in the efforts to address the cosmological
constant problem \cite{Steinhardt}-\cite{Chen}.

The outline of this paper is as follows. In the next section, we
lay the framework for determining the effective 4-dimensional
vacuum energy $\Lambda_4$ by reanalyzing the Randall-Sundrum (RS)
setup \cite{RS1}. In the following section, we start from the
observation that to obtain the critical meV sized $\Lambda_4$
through an exponential suppression from naturally arising vacuum
energies with Planck-scale values, one requires in the exponent a
suppression length directly related to the Grand Unified Theory
(GUT) scale. A geometric brane-world realization of such an
exponential suppression mechanism for $\Lambda_4$ is then given in
terms of two branes embedded in a 5-dimensional ambient spacetime.
The following section analyzes the influence of 5-dimensional bulk
scalars on $\Lambda_4$ by deriving their effective 4-dimensional
potential. In section 5 we embed the 5-dimensional setup into
10-dimensional IIB string-theory using stacks of D3-branes. The
following sections discuss the general features of the
string-theory description. Section 6 investigates how gauge and
Higgs fields emerge from open strings attached to the D3-brane
stacks. A direct consequence of the string-theory description is a
mass hierarchy between color triplets and weak doublets in the
Higgs sector. Section 7 explains how heavy GUT fields together
with light MSSM matter fields can arise from open strings when the
compactification manifold is non-simply connected. Section 8 shows
that the open string spectrum contains the complete spectrum of a
supersymmetric SU(5) GUT theory with gauge group broken down to
the MSSM's $\text{SU(3)}_c \times \text{SU(2)}_L \times
\text{U(1)}_Y$. Finally section 9 addresses the issue of
supersymmetry breaking. In two appendices, we deal with
generalizations of the brane setup to branes with unequal tensions
and analyze the influence of bulk scalars on $\Lambda_4$ for this
unequal tension case.

\section{The Effective Vacuum Energy}

Let us start by analyzing in detail the contributions to the
effective 4-dimensional vacuum energy resp.~cosmological constant
in the RS scenario. The RS-model \cite{RS1} lives in five
dimensions and has two 3-branes with 4-dimensional worldvolume
located at the fixed points of an $\mathbf{S}^1/\mathbf{Z}_2$
orbifold along the fifth direction. In between the 3-branes there
is a bulk 5-dimensional anti de Sitter (AdS) spacetime. The Planck
brane, on which the 4-dimensional graviton gets localized, sits at
the first fixed-point, $x^5=0$ of the $\mathbf{Z}_2$ action,
whereas our 4-dimensional visible world originates from the SM
brane, placed at the second fixed-point $x^5=\pi r$. It is the SM
brane on which the hierarchy problem gets solved by means of the
exponential warp-factor of the AdS bulk geometry. The dominant
contribution to the action of the two branes comes from the brane
tensions $T_{Pl}, T_{SM}$ \cite{Sundrum}. Hence, if one is
interested in a situation where the branes are close to their
ground states, it is reasonable to neglect gauge-field, fermion or
scalar contributions and write for the
RS-Lagrangean\footnote{Subsequently, we will adopt the general
relativity conventions of \cite{Weinberg}.} \cite{RS1}
\begin{alignat}{3}
   S_{RS} = - \int d^4x \int_{0}^{\pi r} dx^5
          &\bigg( \sqrt{-G} \left( M_5^3 R + \Lambda \right) \notag \\
          &+\sqrt{-g_{Pl}^{(4)}} T_{Pl} \delta(x^5)
           +\sqrt{-g_{SM}^{(4)}} T_{SM} \delta(x^5-\pi r)
          \bigg) \; .
\end{alignat}
Here it is understood that we have to integrate the bulk action
over the interval\footnote{This is analogous to the downstairs
approach in heterotic M-theory \cite{HorWitt2}. In the alternative
upstairs approach, one would integrate the Lagrangean density over
the full circle instead but has to add a factor of $1/2$ in front
of the bulk action.} $[-\epsilon,\pi r+\epsilon]$ with $\epsilon$
infinitesimal, rather than $[0,\pi r]$, to incorporate the full
delta-function sources of the boundaries. The 4-dimensional
metrics $g_{SM}^{(4)}, g_{Pl}^{(4)}$ are the respective pullbacks
of the bulk metric to the two 3-brane world-volumes. Adopting the
metric Ansatz
\begin{equation}
   ds^2 = e^{-A(x^5)} \eta_{\mu\nu} dx^\mu dx^\nu + (dx^5)^2 \; ,
    \label{Ansatz}
\end{equation}
the Einstein equations lead to
\begin{equation}
   (A^\prime)^2 = - \frac{1}{3M_5^3} \Lambda \; , \qquad \quad
    A^{\prime\prime} = \frac{1}{3M_5^3}
          \left(T_{Pl} \delta(x^5) + T_{SM} \delta(x^5-\pi r)\right) \; .
    \label{EinsteinEquation}
\end{equation}
The solution to the first equation is given by
\begin{equation}
   A(x^5) = \pm k x^5  \; , \quad k \equiv
            \sqrt{\frac{-\Lambda}{3M_5^3}} \; ,
   \label{BulkSolution}
\end{equation}
where the integration constant has been set to zero. To respect
the orbifold's $\mathbf{Z}_2$ symmetry, which sends
$x^5\rightarrow
-x^5$, we have to take
\begin{equation}
   A(x^5) = \pm k |x^5| \; .
\end{equation}
In the following, we will choose the plus-sign which allows for a
solution of the hierarchy problem on the SM-brane. Noting that
$|x^5|^{\prime\prime}=2\delta(x^5)$, we rewrite the solution in an
expanded form as
\begin{equation}
  A(x^5)= \frac{1}{2}k\left(|x^5|-|x^5-\pi r|\right)
         +\frac{1}{2}k\pi r \; , \qquad
  0 \le x^5 \le \pi r
\end{equation}
in order to satisfy the second equation of
(\ref{EinsteinEquation}) with
\begin{equation}
   T_{Pl}=-T_{SM}=3M_5^3 k \; .
\label{FineTune}
\end{equation}

Let us now determine the resulting 4-dimensional effective action
by integrating out the fifth dimension. We will first carry this
out for the Einstein-Hilbert term of the bulk action. For this
purpose, consider first the general $D$-dimensional case with
warped metric
\begin{alignat}{3}
     ds^2 &= G_{MN} dx^M dx^N
          &= g_{\mu\nu}^{(D-1)} dx^\mu dx^\nu + (dx^D)^2
           = f(x^D) g_{\mu\nu}(x^\rho)\, dx^\mu dx^\nu
             + (dx^D)^2 \; ,
\end{alignat}
where $\mu,\nu$ run over $1,\hdots,D-1$ and $M,N$ over
$1,\hdots,D$. The $D$-dimensional curvature scalar can then be
decomposed as follows into the $(D-1)$-dimensional curvature
scalar plus additional terms depending exclusively on $x^D$
($f^\prime$ denotes the derivative $df/dx^D$)
\begin{equation}
     R(G) = \frac{1}{f} R(g) + \frac{1}{4} (D-1)
                 \left( (D-2) \big[ (\ln f)^\prime \big]^2
                       +2(\ln f)^{\prime\prime} + 2 \frac{f^{\prime\prime}}{f}
                 \right) \; .
        \label{RiemannScalar}
\end{equation}
In addition, we have to take into account a factor $\sqrt{-G} =
f^{(D-1)/2}\sqrt{-g}$ in the measure of the action integral.

Specializing now to the RS case with $D=5$ we have the metric
\begin{equation}
ds^2 = G_{MN} dx^M dx^N
     = e^{-A(x^5)} g_{\mu\nu} (x^\rho) dx^\mu dx^\nu + (dx^5)^2 \; .
   \label{Metric}
\end{equation}
Using (\ref{RiemannScalar}) with $f(x^5) = e^{-A(x^5)}$, we get
\begin{alignat}{3}
   S_{\text{EH}} &= -\int d^4x \int_0^{\pi r} dx^5 \sqrt{-G} M_5^3 R(G) \notag \\
                 &= - \int d^4x \sqrt{-g} M_5^3
                      \int_0^{\pi r} dx^5
               \left( e^{-A} R(g)
                     + e^{-2A} \left[5(A^\prime)^2-4A^{\prime\prime}\right]
               \right) \; .
       \label{EHReduction}
\end{alignat}
Since we will make use of this formula later on, we note that up
to this point it is valid for any metric which is of the form
(\ref{Metric}). Choosing the RS-metric we obtain
\begin{alignat}{3}
    S_{\text{EH}} = - \int d^4x \sqrt{-g} M_5^3
             &\bigg( R(g) \int_0^{\pi r} dx^5 e^{-kx^5} \notag
              \\
              &+ \int_0^{\pi r} dx^5 e^{-2kx^5}
              \big[ 5k^2
                   -4k \left( \delta(x^5) - \delta(x^5-\pi r)
                       \right)
              \big]
              \bigg)  \; .
\end{alignat}
Concerning the delta-function integration we perform the
integration over the interval $[-\epsilon, \pi r+\epsilon]$ with
$\epsilon$ infinitesimal. Thus the Einstein-Hilbert action
contributes the terms
\begin{equation}
  S_{\text{EH}}
  = - \int d^4x \sqrt{-g}
             \left( M_4^2 R(g)
                    -\frac{3}{2}M_5^3 k \left(1-e^{-2k\pi r}\right)
             \right) \; ,
\end{equation}
to the 4-dimensional effective action, where $M_4^2=2 M_5^3 (1
- e^{-k\pi r})/k$ denotes the 4-dimensional Planck-scale squared.

The second part of the reduction comprises the brane sources and
the bulk cosmological constant term
\begin{alignat*}{3}
   S_{Pl}+S_{SM}+S_\Lambda
                       &= - \int d^4x \sqrt{-g}
                            \left( e^{-2k\pi r} T_{SM} + T_{Pl}
                                    + \Lambda\int_0^{\pi r} dx^5 e^{-2kx^5}
                            \right) \\
                       &= -\frac{3}{2} \int d^4x \sqrt{-g} M_5^3 k
                           \left(1-e^{-2k\pi r}\right) \; ,
\end{alignat*}
where we have used (\ref{FineTune}) in the last row. To obtain the
effective potential in the RS-scenario we add both contributions.
Because $R(g)$ vanishes and due to the fine-tuning of the Planck
and SM brane tensions in terms of the bulk cosmological constant
(\ref{FineTune}), both contributions add up to zero and we obtain
a zero $\Lambda_4$, as expected.

There are two interesting observations at this point. First, the
above calculation shows that if one relaxes the fine-tuning
(\ref{FineTune}) of the brane tensions but still assumes that they
are equal in magnitude and of opposite sign, then one expects a
residual 4-dimensional vacuum energy of order
\begin{equation}
   \Lambda_4 \approx \pm M_5^3 k (1-e^{-2k\pi r}) \; .
   \label{RSCoCo}
\end{equation}
where the sign depends on whether the bulk cosmological constant
$\Lambda$ is larger than the brane tensions (minus sign) or
smaller (plus sign). Such an effective $\Lambda_4$ would
constitute a potential for the interval length modulus $r$. For
the plus sign choice its minimum lies at $r=0$ and would drive
$\Lambda_4$ to zero\footnote{Note that in this limit the
hierarchy-problem couldn't be solved any longer.} (for the minus
sign choice the minimum would lie at $r=\infty$ implying a runaway
behavior). However, an estimate of how close $r$ has to come to
zero to reconcile the vacuum energy with its observable value is
rather disenchanting. If we take $M_5^3 k \approx M_{Pl}^4$,
$k\approx M_{Pl}$ and demand that $\Lambda_4\approx
\text{1meV}^4$, we find an incredibly small $r\approx 10^{-125}
l_{Pl}$, where $l_{Pl}=M_{Pl}^{-1}$ with the Planck-mass given by
$M_{Pl}=1.2\times 10^{19}\,$GeV. This, however, is a region, where
we surely cannot trust classical gravity any longer as a reliable
framework.

Second, one observes that the warp-factor enters $M_4^2$ and
$\Lambda_4$ differently. This means that the exponential
warp-factor contribution to the cosmological constant $\lambda_4 =
\Lambda_4/M_4^2$ does not drop out and therefore presents an
interesting possibility to influence the effective 4-dimensional
cosmological constant if one could get rid of the constant
$r$-independent terms which exceed the exponential terms. In the
rest of this paper we describe a 5-dimensional brane-world
scenario where we study this warp-factor influence on $\lambda_4$
or equivalently the vacuum energy $\Lambda_4$ and its implications
for string- and particle-theory.

\section{The 5-Dimensional Two Brane Model}

In the previous section we saw that the asymmetric (with respect
to the $\mathbf{Z}_2$ orbifold symmetry acting on the fifth
dimension) positioning of the Planck brane at $x^5=0$ and the SM
brane at $x^5=\pi r$ led, together with the choice of asymmetric
tensions for these two branes, to an asymmetric warp-factor. It is
this asymmetry which generated the unwanted constant term (from
the Planck brane) next to the $r$-dependent wanted exponential
term (from the SM brane) in the 4-dimensional vacuum energy
(\ref{RSCoCo}). To obtain a small $\Lambda_4$ one should hence
avoid placing a brane at the origin, the fixed point of the
$\mathbf{Z}_2$ symmetry. Instead, we will place the two branes at
the $\mathbf{Z}_2$ mirror-symmetric points $x^5=-l$ and $x^5=l$.
We will see that in this way one can achieve an exponential
suppression of $\Lambda_4$ at the expense of one classical
fine-tuning. In contrast to the RS-model we will take $x^5$ to be
non-compact, much as in the second model of Randall and Sundrum
\cite{RS2}. To respect the $\mathbf{Z}_2$ symmetry both branes
will be given the same positive tension $T$ (see
fig.~\ref{PictureTwoWalls}). Since we are assuming a non-compact
$x^5$ coordinate, it is consistent to have both tensions
positive\footnote{Only for closed, i.e.~compact and without
boundary, dimension $x^5$ can one show that the sum of all brane
tensions has to vanish \cite{GKL}.}. In the bulk we will adopt a
piecewise constant cosmological constant $\Lambda(x^5)$ so that
the complete 5-dimensional action becomes
            \setcounter{figure}{0}
            \begin{figure}[t]
              \begin{center}
               \begin{picture}(150,120)(-10,0)
                  \Line(15,0)(15,24)
                  \Line(15,45)(15,100)
                  \Line(15,100)(25,120)
                  \Line(15,0)(25,20)
                  \Line(25,20)(25,35)
                  \Line(25,45)(25,120)

                  \Line(85,0)(85,24)
                  \Line(85,45)(85,100)
                  \Line(85,100)(95,120)
                  \Line(85,0)(95,20)
                  \Line(95,20)(95,35)
                  \Line(95,45)(95,120)

                  \LongArrow(-30,40)(140,40)
                  \Text(153,43)[]{$x^5$}
                  \Text(10,33)[]{$-l$}
                  \Text(50,34)[]{$0$}
                  \Line(50,39)(50,41)
                  \Text(85,33)[]{$l$}
                  \Text(126,77)[]{$\text{AdS}_5$}
                  \Text(126,60)[]{Half-Slice}
                  \Text(-15,77)[]{$\text{AdS}_5$}
                  \Text(-15,60)[]{Half-Slice}
               \end{picture}
               \caption{The double 3-brane setup with interbrane separation $2l$.
                        To obtain a critical vacuum energy one has
                        to set $2l\approx 1/M_{\text{GUT}}$.}
               \label{PictureTwoWalls}
              \end{center}
            \end{figure}
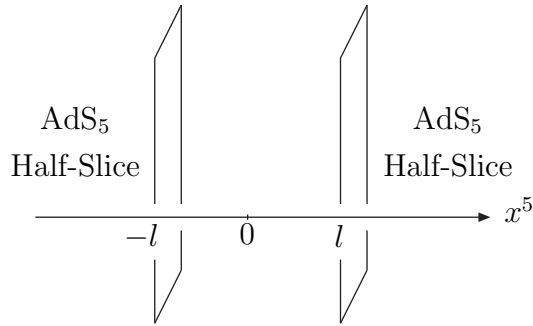
\hspace{-2.5mm}
\begin{alignat}{3}
S&=-\int d^4x\int dx^5 \sqrt{-G}\left( M_5^3 R(G)+\Lambda \right) \notag \\
   &\phantom{=}
     -\int d^4x\int dx^5 \left( \sqrt{-g^{(4)}_1} T_1
                                \delta\left(x^5+l\right)
                               +\sqrt{-g^{(4)}_2} T_2
                                \delta\left(x^5-l\right)
                         \right) \; .
\end{alignat}
where we take $T_1=T_2=T$ and relegate the case with unequal
tensions to appendix A. Again $g_{1,\mu\nu}^{(4)}$ and
$g_{2,\mu\nu}^{(4)}$ are the induced metrics arising from the
pullback of the bulk metric $G_{MN}$ to the two brane
world-volumes.

\begin{figure}[ht]
              \begin{center}
               \begin{picture}(150,140)(-10,0)
                  \LongArrow(50,9)(50,90)
                  \Text(50,104)[]{$A(x^5)$}

                  \LongArrow(-20,10)(130,10)
                  \Text(142,13)[]{$x^5$}

                  \Text(0,3)[]{$-l$}
                  \Line(0,9)(0,11)

                  \Text(50,3)[]{$0$}

                  \Text(100,3)[]{$l$}
                  \Line(100,9)(100,11)

                  \Line(-10,100)(0,38)
                  \Line(0,38)(100,38)
                  \Line(100,38)(110,100)

                  \Text(38,48)[]{$k l$}
               \end{picture}
               \caption{The function $A(x^5)$ which determines the
                        warp-factor along the non-compact fifth dimension.}
               \label{PictureWarpSimple}
              \end{center}
            \end{figure}
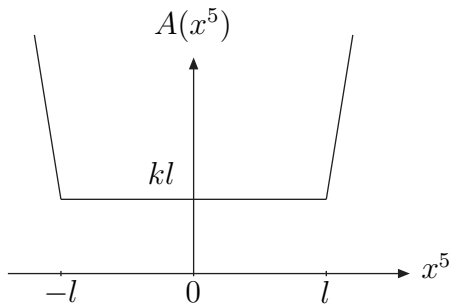

Choosing once more the Ansatz
\begin{equation}
ds^2 = e^{-A(x^5)} \eta_{\mu\nu} dx^\mu dx^\nu + (dx^5)^2 \; ,
\label{Solution}
\end{equation}
the Einstein field equations reduce to the two equations
(\ref{EinsteinEquation}) which read in our case
\begin{equation}
(A^\prime)^2 = - \frac{1}{3M_5^3} \Lambda(x^5) \; , \qquad \quad
A^{\prime\prime} = \frac{1}{3M_5^3}
\left(T \delta\left(x^5+l\right)
     +T \delta\left(x^5-l\right)\right) \; .
    \label{Sources}
\end{equation}
The solution to these equations is given by
\begin{equation}
   A(x^5) \; = \; \frac{k}{2}\left|x^5+l\right|
           +\frac{k}{2}\left|x^5-l\right|
          \; = \;
            \left\{ \begin{array}{cl}
      -kx^5 &,\,\,  x^5 \le -l \\
      kl &,\,\,  -l \le x^5 \le l \\
      k x^5 &,\,\,  x^5 \ge l
                    \end{array}
            \right.
       \; ,
    \label{WarpFactor}
\end{equation}
together with a bulk cosmological constant
\begin{equation}
   \Lambda(x^5) \; = \;
               \left\{ \begin{array}{cl}
             \Lambda  &,\,\,  |x^5| > l\\
             \Lambda/4  &,\,\,  |x^5| = l\\
             0  &,\,\,  |x^5| < l
                       \end{array}
               \right.
                \; = \;
               \left\{ \begin{array}{cl}
             -3M_5^3k^2  &, \,\, |x^5| > l\\
             -3M_5^3k^2/4  &, \,\, |x^5| = l\\
                 0     &, \,\, |x^5| < l
                       \end{array}
               \right.
        \label{LambdaSimple}
\end{equation}
and brane-tension
\begin{equation}
    T = 3M_5^3 k \; .
        \label{TensionSimple}
\end{equation}
Here we have set the integration constant, which could have been
added to $A(x^5)$, to zero which amounts to one fine-tuning at the
classical level\footnote{The same solution but with another choice
for the undetermined integration constant has been obtained in
\cite{Lykken}. The analysis in this work focussed on the
localization of the graviton and corrections to the Newtonian
gravitational potential.}. The relation between the bulk
cosmological constant in the exterior $\ads_5$ half-patches and
the 3-brane tension becomes
\begin{equation}
  \Lambda = -\frac{1}{3}\frac{T^2}{M_5^3} \; .
   \label{FineTunedParameters}
\end{equation}
The $\mathbf{Z}_2$ symmetric function $A(x^5)$ which determines
the warp-factor is displayed in fig.\ref{PictureWarpSimple}.
Notice that the warp-factor is smaller than $e^{-k l}$ everywhere
\begin{equation}
e^{-A(x^5)} \le e^{-k l} \; .
\end{equation}
For a low-energy observer at energies below $1/2l$, which we will
soon identify with the GUT scale, the separation between the two
3-branes becomes invisible. As a result he will see a geometry of
two slices of $\ads_5$ spacetime glued together. For him the
graviton would hence appear localized on the merged two 3-branes
as described in \cite{RS2}.

Next, let us derive the effective 4-dimensional action by
integrating over the $x^5$ coordinate in the action. For this we
adopt the slightly more general background
\begin{equation}
  ds^2 = e^{-A(x^5)}g_{\mu\nu}(x^\rho)dx^\mu dx^\nu+(dx^5)^2 \; .
  \label{GeneralMetric}
\end{equation}
where the flat $\eta_{\mu\nu}$ is replaced by a general
$g_{\mu\nu}$, thus also allowing for 4-dimensional spacetimes with
non-vanishing $\Lambda_4$. Along the same lines as for the RS-case
we obtain by using (\ref{EHReduction}) for the Einstein-Hilbert
term
\begin{alignat}{3}
   S_{EH} &= -\int d^4 x \sqrt{g} M_5^3
             \left( R(g) \int_{-\infty}^\infty dx^5
             e^{-A} + \int_{-\infty}^\infty dx^5 e^{-2A}
             \left[ 5(A^\prime)^2-4A^{\prime\prime} \right]
                                      \right) \notag \\
          &= -e^{-kl} \int d^4x \sqrt{g}M_5^3
            \left( 2R(g) \Big[\frac{1}{k}+l\Big]
             - 3 e^{-kl} k \right) \; .
\end{alignat}
Combining the two brane actions and the bulk cosmological constant
gives furthermore
\begin{equation}
   S_{SM_1}+S_{SM_2}+S_\Lambda = -e^{-2kl} \int d^4x \sqrt{g}
         \left( 2T+\frac{\Lambda}{k} \right) \; .
\end{equation}
Taken together the total 4-dimensional effective action becomes
\begin{alignat}{3}
   &\phantom{=\;\;\,}S_{EH}+S_{SM_1}+S_{SM_2}+S_\Lambda \notag \\
          &= -e^{-kl}\int d^4x\sqrt{g}
   \left( 2M_5^3 R(g)\Big[\frac{1}{k} + l\Big]
             + e^{-kl}\Big[ -3M_5^3k+2T+\frac{\Lambda}{k} \Big]
   \right) \; .
\end{alignat}
We will now drop the overall constant scale-factor $e^{-kl}$ since
it drops out of the equations of motion. One can also show by
replacing $(dx^5)^2$ in the Ansatz for the metric by the more
general $e^{-B(x^5)}(dx^5)^2$ that this constant overall factor
can be absorbed without loss of generality into the definition of
$x^5$. Our final 4-dimensional effective action therefore reads
\begin{equation}
       S_{D=4}  = -\int d^4x\sqrt{g}
   \left( M_4^2 R(g) + \Lambda_4 \right) \; ,
             \label{FourAction}
\end{equation}
with the effective Planck-scale $M_4$ and vacuum energy
$\Lambda_4$ given by
\begin{alignat}{3}
   M_4^2 &= 2M_5^3 \Big(\frac{1}{k}+l\Big)
       \label{EffectiveMass} \\
   \Lambda_4 &= e^{-kl}\Big(-3M_5^3 k+2T+\frac{\Lambda}{k}\Big) \; .
       \label{EffectiveCoCo}
\end{alignat}
The remaining exponential factor $e^{-kl}$ which occurs only in
the vacuum energy will play an important role soon. But before
coming to that, let us quickly verify our result by plugging in
the values (\ref{LambdaSimple}), (\ref{TensionSimple}) for
$T,\Lambda$ of our solution (\ref{WarpFactor}) which guaranteed a
flat 4-dimensional Minkowski background. Therefore, thanks to the
tuning of these parameters expressed by the relations
(\ref{TensionSimple}) and (\ref{FineTunedParameters}), the
4-dimensional vacuum energy $\Lambda_4$ must vanish. This is
indeed what we find with the above expression for $\Lambda_4$ and
serves as a check on its derivation.

The important point is however the following. Suppose we lift the
finetuning imposed on the parameters $\Lambda$ and $T$. The
4-dimensional metric would then be no longer flat and the
background becomes\footnote{It has been shown in \cite{AK4} that
the full backreaction of the non-finetuned parameters preserves
this warp-factor structure of the metric.}
\begin{equation}
  ds^2 = e^{-A(x^5)}g_{\mu\nu}dx^\mu dx^\nu + (dx^5)^2 \; .
\end{equation}
Without tuning, the square bracket in (\ref{EffectiveCoCo}) will
generically assumes positive or negative values of order
$M_{Pl}^4$. In this paper we want to focus on the positive values.
Taking the fundamental 5-dimensional Planck scale at $M_5 =
M_{Pl}$, it will be natural to have also $k = M_{Pl}$. Note that
the bulk cosmological constant will stay zero in between the
3-branes if we leave the $\mathbf{Z}_2$ symmetry of the setup
intact \cite{AK4}. The 4-dimensional vacuum energy for non-tuned
parameters will then be non-zero and of the form
\begin{equation}
\Lambda_4 = e^{-k l} M_{Pl}^4 \; .
\label{PlanckGUTRelation1}
\end{equation}
This is an interesting result since the exponential factor allows
to lower the enormous Planck sized vacuum energy down to small
values in a natural way, i.e.~without invoking a new large
hierarchy. So what is the appropriate distance $2l$ between the
two 3-branes which allows to bring a Planck sized vacuum energy
down to the critical meV scale? With $k = M_{Pl}$ this length
coincides quite precisely with the inverse GUT unification scale
\begin{equation}
  \begin{array}{|c|}
  \hline \\[-5mm]
  \;\; \Lambda_4 \approx \text{meV}^4
  \qquad\Leftrightarrow\qquad
  2l \approx M_{\text{GUT}}^{-1} \;\; \\[1mm]
  \hline
  \end{array}
    \label{PlanckGUTRelation2}
\end{equation}
which is given by $M_{\text{GUT}} = 2 \times 10^{16}\,\text{GeV}$
or $M_{\text{GUT}} = (568l_{Pl})^{-1}$ in terms of the Planck
length\footnote{Notice that because $2l\gg l_{Pl}$ we can trust
the field theory framework.}. It is intriguing that the required
length turns out to be so natural which strongly suggests some GUT
theory connection. We will make this connection explicit later
when we embed the 5-dimensional setup into 10-dimensional IIB
string-theory.

Let us take stock of what has been achieved so far. Using one
classical tuning to set the integration constant in
(\ref{WarpFactor}) to zero (one might hope to find a dynamical
reason for this natural choice), we obtain with the warped
geometry a mechanism to exponentially suppress the generically
Planck sized vacuum energy down to critical $\text{meV}^4$ size.
In particular this allows to suppress quantum corrections to the
4-dimensional vacuum energy coming from fields on the 3-branes,
which renormalize the brane's tension $T$, without the need to
readjust the resulting vacuum energy order by order in
perturbation theory. Besides suppressing the contributions to the
4-dimensional vacuum energy coming from classical bulk
contributions and classical plus quantum contributions of gauge
and matter fields located on the 3-branes, it would be very
interesting to investigate whether the suppression mechanism also
extends to quantum contributions coming from bulk fields. We will
leave this interesting aspect to future research but verify as a
first step in this direction in the next section that generic bulk
fields do not spoil the suppression mechanism at the classical
level. The intriguing outcome of this section is furthermore that
this warped geometry vacuum energy suppression mechanism points
directly towards a connection between the $\text{meV}^4$ critical
vacuum energy $\Lambda_4$, the Planck scale $M_{Pl}$ and the GUT
unification scale $M_{GUT}$. The role of GUT theories will be
explored in sections 5 to 8.

\section{The Effective Potential from Bulk Scalars}

An embedding of the 5-dimensional setup into IIB string-theory or
F-theory along the lines of \cite{CPV} requires in general
additional bulk scalar fields in the 5-dimensional theory coming
from the decomposition of 10-dimensional fields upon dimensional
reduction from ten to five dimensions. It is important to check
that such additional bulk fields do not reintroduce unsuppressed
Planck-scale contributions to the 4-dimensional vacuum energy.
Otherwise the warp-factor suppression mechanism of the
4-dimensional vacuum energy described so far could only be
realized in simple 5-dimensional brane world scenarios without
bulk scalars but not be embedded into string-theory. We will show
in this section that generic bulk scalars do not spoil the
suppression mechanism.

For this let us now examine the 4-dimensional effective potential,
in the same warped background as before, generated by a canonical
5-dimensional bulk scalar $\Phi$ with quartic couplings to the two
3-branes. Such couplings are for instance required by the
Goldberger-Wise mechanism \cite{GoldWise} to stabilize the fifth
dimension. The action for this scalar reads
\begin{alignat}{3}
   S_\Phi =& -\int d^4x \int_{-\infty}^\infty dx^5 \sqrt{G}
             \bigg( \frac{1}{2}G^{MN}\partial_M\Phi\partial_N\Phi
                    +\frac{1}{2}m^2\Phi^2
             \bigg)      \notag \\
           & -\int d^4x \int_{-\infty}^\infty dx^5
             \bigg( \sqrt{g_1^{(4)}} \lambda_1 (\Phi^2-v_1^2)^2\delta(x^5+l)
                    +\sqrt{g_2^{(4)}} \lambda_2 (\Phi^2-v_2^2)^2\delta(x^5-l)
             \bigg)      \notag \; ,
\end{alignat}
with $m$ the scalar's mass and positive couplings $\lambda_1,
\lambda_2$. We will assume that $\Phi$ depends only on $x^5$ and
study it in the fixed gravitational background given by our
solution (\ref{Solution}), (\ref{WarpFactor}). In this background
we arrive at the following equation of motion for $\Phi$
\begin{alignat}{3}
   ( e^{-2A(x^5)}\Phi^\prime )^\prime - e^{-2A(x^5)}m^2\Phi
  = 4 \big[&e^{-2A(-l)}\lambda_1 (\Phi^2-v_1^2) \Phi \delta(x^5+l) \notag \\
          +&e^{-2A(l)}\lambda_2 (\Phi^2-v_2^2) \Phi \delta(x^5-l)
      \big] \; ,
\end{alignat}
which has the solution
\begin{equation}
   \Phi(x^5)= \left\{
                   \begin{array}{cl}
                      a e^{(1+\Gamma)A(x^5)} + b e^{(1-\Gamma)A(x^5)}
                     &,\,\, x^5  < -l \;  \\
                      c e^{mx^5} + d e^{-mx^5}
                     &,\,\,|x^5| \le l \\
                      e e^{(1+\Gamma)A(x^5)} + f e^{(1-\Gamma)A(x^5)}
                     &,\,\, x^5 > l
                   \end{array}
              \right.\; ,
\end{equation}
with
\begin{equation}
  \Gamma = \sqrt{1+m^2/k^2}
\end{equation}
and free coefficients $a,b,c,d,e,f$.

In order to obtain a normalizable solution we are forced to set
$a=e=0$. Furthermore, demanding continuity of $\Phi$ at the
location of the 3-branes determines $b$ and $f$ in terms of $c$
and $d$
\begin{alignat}{3}
  b&= e^{(\Gamma-1)kl} \btil \; , \quad
  \btil = c e^{-ml} + d e^{ml}    \\
  f&= e^{(\Gamma-1)kl} \ftil \; , \quad
  \ftil = c e^{ml} + d e^{-ml} \; .
\end{alignat}
To fix the remaining free coefficients $c$ and $d$ one could plug
the above solution into the field equation and integrate it over
the fifth dimension to incorporate brane boundary conditions.
However, this leads to a complicated cubic equation in the
unknowns $c$ and $d$. An easier way to arrive at a determination
of the coefficients $c$ and $d$, proposed by \cite{GoldWise}, is
to insert the scalar field solution into the scalar's action and
integrate over $x^5$ to arrive at an effective potential for the
interbrane distance $2l$. The minimization of this effective
potential will then determine $c$ and $d$. We will now follow this
strategy. From the couplings of $\Phi$ to the 3-branes the
effective potential receives the contributions
\begin{equation}
  \int d^4x \left( \sqrt{g_1^{(4)}}\lambda_1\left(\Phi^2(-l)-v_1^2\right)^2
                   +\sqrt{g_2^{(4)}}\lambda_2\left(\Phi^2(l)-v_2^2\right)^2
            \right) \; .
\end{equation}
To minimize this potential for positive couplings
$\lambda_1,\lambda_2$ we must set $\Phi(-l)=v_1$ and
$\Phi(l)=v_2$. These two conditions finally determine $c$ and $d$
as
\begin{equation}
   c=\frac{-v_1 e^{-ml} + v_2 e^{ml}}{2\sinh(2ml)} \; , \qquad
   d=\frac{v_1 e^{ml} - v_2 e^{-ml}}
          {2\sinh(2ml)} \; .
\end{equation}

With all coefficients in the solution for $\Phi$ being fixed, the
effective 4-dimensional potential $V_\Phi$, defined by
$S_\Phi=-\int d^4x \sqrt{g} V_\Phi(l)$, becomes
\begin{equation}
   V_\Phi(l) = \frac{e^{-2kl}}{2}
               \left( (v_1^2+v_2^2)
                       \left[ (\Gamma-1)k+m\coth(2ml) \right]
                      - 2v_1v_2\frac{m}{\sinh(2ml)}
               \right) \; ,
   \label{FullPotential}
\end{equation}
where the identity
\begin{equation}
(1-\Gamma)^2 k^2+m^2 = 2\Gamma(\Gamma-1)k^2
\end{equation}
has been utilized. Usually when performing a dimensional reduction
of a string-theory model, we retain only the massless modes with
$m=0$ in the low-energy effective action. For these the effective
potential generated by $\Phi$ simplifies to
\begin{equation}
   V_\Phi(l) = \frac{e^{-2kl}}{4l}(v_1-v_2)^2 \; .
   \label{SimplePotential}
\end{equation}
Therefore both in the massive and massless case, the important
exponential suppression-factor is present (again only one
$e^{-kl}$ remains after discarding an overall $e^{-kl}$ factor of
the action as explained earlier). Hence, for the same distance
between the 3-branes as before, $2l = M^{-1}_{\text{GUT}}$, bulk
scalars with values for $v_1, v_2, m$ up to the Planck scale will
not introduce contributions to the 4-dimensional vacuum energy
larger than the critical one in virtue of
(\ref{PlanckGUTRelation1}), (\ref{PlanckGUTRelation2}).

As an aside, let us ask whether the effective potential obtained
from a bulk scalar may stabilize the interbrane distance $2l$.
From (\ref{SimplePotential}) it is immediately recognizable that
in the massless case no minimum at finite $l$ exists. In the
massive case, setting $\partial V_\Phi/\partial l$ equal to zero,
leads to the following equation for $l$
\begin{equation}
   (w^2+1)\left(\left[\frac{\Gamma-1}{r}+\cosh(2ml)\right]\sinh(2ml)+r\right)
  =2w\left(r\cosh(2ml)+\sinh(2ml)\right) \; ,
\end{equation}
where we have employed the dimensionless ratios
\begin{equation}
  w = \frac{v_1}{v_2} \;, \qquad r = \frac{m}{k} \; ,
\end{equation}
in terms of which we can write $\Gamma=\sqrt{1+r^2}$. A numerical
analysis of this equation for generic values of $v_1, v_2, m$
shows that there are no solutions for $l$ which would be real and
positive. We can therefore conclude that in the massive case the
effective potential exhibits no minimum either. Thus bulk scalars
cannot be used for a stabilization of $l$. The case with different
3-brane tensions $T_1\not= T_2$ will be analyzed in appendix B.
Let us note that for the IIB string-theory embeddings which we
will discuss in the next section following the uplifting of
RS-scenarios as described in \cite{CPV}, only scalars are
generated in five dimensions upon dimensional reduction of the
internal metric and other fields including the axio-dilaton,
3-form and 5-form fluxes.

The inability of the scalars to stabilize the fifth dimension open
up, however, a potential relevance for cosmology. For $2l =
M_{\text{GUT}}^{-1}$ we find that bulk scalars induce an extremely
tiny (since exponentially suppressed) but non-vanishing repulsive
force between the 3-branes such that the setup might be regarded
as quasi-static. On the other hand for much smaller lengths
$2l\approx 0$ two things will happen. First, the repulsive force
will be much larger, driving the two 3-branes apart very quickly,
hence leading to a fast time-dependent cosmological evolution.
Second, the vacuum energy will no longer be suppressed as the
exponential factor becomes unity. This seems to fit well with
expectations about the very early universe, where to start
inflation a considerable nonvanishing 4-dimensional vacuum energy
$\Lambda_4 = V(\phi)$ is needed, $V(\phi)$ being the potential or
vacuum energy density of the inflaton $\phi$. For example in the
scenario of chaotic inflation \cite{Linde} one indeed requires a
Planck size potential $V(\phi)\approx M_{Pl}^4$ which could arise
when $2l\approx 0$. And there is another aspect which fits nicely
in. Namely at $2l\approx 0$ both 3-branes lie essentially on top
of each other. If we jump a bit ahead and identify the 3-branes
with D3-branes in IIB string-theory, then we know that putting
them on top of each other implies a gauge symmetry enhancement
which, as will be discussed later, could describe a GUT
unification. This then suggests the following cosmological
scenario. In the very early universe when both 3-branes are close
together we have an unbroken GUT unification group and a huge
cosmological constant, potentially capable of driving inflation.
Due to the large repulsive force between the 3-branes they
initially separate rapidly along the fifth direction. However, the
separation process slows down soon due to the exponential
suppression of the repulsive interbrane force. Today these forces
have become miniscule and the brane setup evolution quasi-static
with interbrane distance $2l$ having reached $M_{\text{GUT}}^{-1}$
giving a small critical vacuum energy. Moreover, the GUT
unification group will be broken today upon identification of the
3-branes with D-branes. An evolution along these lines might also
arise in heterotic M-theory where forces between its two
boundaries depend similarly on their distance \cite{KraHet}. We
will not investigate these cosmological aspects further in this
work and will now discuss the string-theory embedding.

\section{Lift to IIB String-Theory}

We have seen the important role played by the GUT unification
scale in the suppression of the vacuum energy to achieve the
critical value. It arose geometrically as the inverse of the
length between the two 3-branes and strongly suggests a GUT theory
connection. This connection and the GUT theory will become
transparent once the 5-dimensional setup is embedded into IIB
string-theory resp.~F-theory compactified on a Calabi-Yau
three-fold resp.~four-fold. The 3-branes become D3-branes and open
strings stretching between them over a length of the inverse GUT
scale give naturally rise to $X, Y$ leptoquark gauge bosons with
masses at or above $M_{\text{GUT}}$. In consistency with the fact
that we are addressing the vacuum energy not at early epochs of
the universe but today, we will find a GUT theory with broken
symmetry and consequently heavy $X$ and $Y$ leptoquark gauge
bosons. We will discuss the string-theory GUT connection and
related issues in this and the remaining sections.

\begin{figure}[t]
\begin{center} \begin{picture}(150,160)(-10,-30) \Line(10,0)(10,24)
\Line(10,45)(10,100) \Line(10,100)(20,120) \Line(10,0)(15,10)
\Line(20,20)(20,24) \Line(20,110)(20,120)

                  \Line(15,0)(15,24)
                  \Line(15,45)(15,100)
                  \Line(15,100)(25,120)
                  \Line(15,0)(20,10)
                  \Line(25,110)(25,120)

                  \Line(20,0)(20,24)
                  \Line(20,45)(20,100)
                  \Line(20,100)(30,120)
                  \Line(20,0)(30,20)
                  \Line(30,20)(30,35)
                  \Line(30,45)(30,120)

                  \Text(15,-10)[]{\footnotesize SU(3)}

                  \Line(77,0)(77,35)
                  \Line(77,45)(77,100)
                  \Line(77,100)(87,120)
                  \Line(77,0)(87,20)
                  \Line(87,20)(87,24)
                  \Line(87,45)(87,120)

                  \Line(92,0)(92,24)
                  \Line(92,45)(92,100)
                  \Line(92,100)(102,120)
                  \Line(92,0)(97,10)
                  \Line(102,110)(102,120)

                  \Line(97,0)(97,35)
                  \Line(97,45)(97,100)
                  \Line(97,100)(107,120)
                  \Line(97,0)(107,20)
                  \Line(107,20)(107,38)
                  \Line(107,45)(107,120)

                  \Text(75,-10)[]{\footnotesize U(1)}
                  \Text(102,-10)[]{\footnotesize SU(2)}

                  \LongArrow(-20,40)(130,40)
                  \Text(143,42)[]{$X^5$}
                  \Text(11,33)[]{$-l_\sigma$}
                  \Text(50,32)[]{$0$}
                  \Line(50,39)(50,41)
                  \Text(89,33)[]{$l_\sigma$}

                  \Text(158,60)[]{$\text{AdS}_5$ throat}
                  \DashCurve{(120,120)(160,100)(230,80)}{5}
                  \DashCurve{(120,0)(160,20)(230,40)}{5}
                  \Text(-40,60)[]{$\text{AdS}_5$ throat}
                  \DashCurve{(0,120)(-40,100)(-110,80)}{5}
                  \DashCurve{(0,0)(-40,20)(-110,40)}{5}

               \end{picture}
               \caption{The two 3-branes resolved as stacks of D3-branes in
                        the microscopic IIB string-theory picture.}
               \label{PictureBranes}
              \end{center}
            \end{figure}
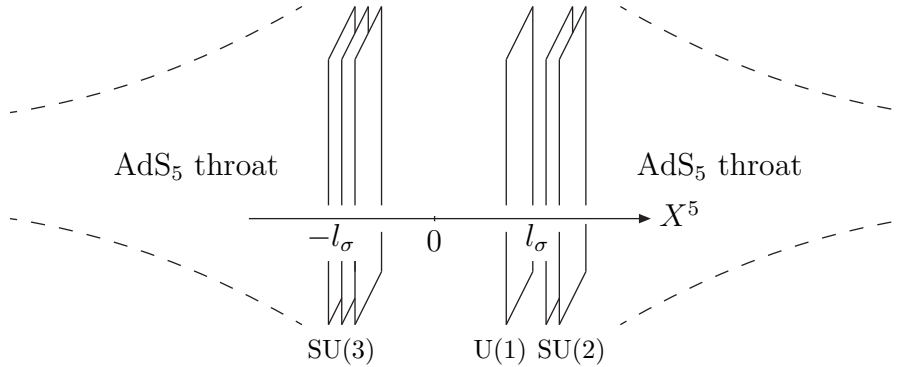

The low-energy 5-dimensional geometry consists of two half
infinite $\ads_5$ patches with an interpolating flat spacetime
interval. Since an $\ads_5$ geometry arises as the near-horizon
geometry of a type IIB string-theory D3-brane, one is naturally
led to consider IIB string-theory as the 10-dimensional parent
theory and replace the 3-branes by D3-branes. Moreover, since we
had two 3-branes of equal tension, we should consider two stacks
of D3-branes with the same amount of D3-branes in either stack.
Now we have to decide how many D3-branes should be in each stack.
For this let us note that we also need to accomodate the SM with
gauge group $[\text{SU(3)}_c] \times [\text{SU(2)}_L \times
\text{U(1)}_Y]$. With two stacks of an equal number of D3-branes
this can be achieved by having three D3-branes in each stack. The
QCD color group will arise from one stack while the electroweak
gauge group arises from the other. The split into the weak
$\text{SU(2)}_L$ and electromagnetic $\text{U(1)}_Y$ gauge group
requires a tiny further split between the 3 branes of the second
stack into 2 giving rise to $\text{SU(2)}_L$ and a single one
responsible for $\text{U(1)}_Y$ (see fig.\ref{PictureBranes}).
By $X^5$ we will denote the string-theory target space coordinate
which relates upon reduction to five dimensions to $x^5$. Each
stack of D3-branes gives rise to a further local U(1) symmetry
which is related to the center of mass position of the stack. For
these two and the $\text{U(1)}_Y$ gauge group it can be shown that
only one can stay anomaly-free \cite{AIQU}, \cite{AKT}.
$\text{U(1)}_Y$ will later be identified with the anomaly-free
hypercharge gauge group since the SM matter fields will turn to
transform under it with the correct hypercharges. The anomalies of
the two other abelian U(1) groups are cancelled in string-theory
by the Green-Schwarz mechanism, which renders them massive. They
remain as global symmetries with the mass of the corresponding
U(1) gauge-bosons shifted to the string-scale.

This general recipe for lifting 5-dimensional $\ads_5$ geometries
to 10-dimensional IIB string-theory setups containing D3-branes
whose near-horizon geometries give rise to the $\ads_5$ throats
has been proposed in \cite{CPV} and will be used here. Indeed for
the $\mathbf{Z}_2$ symmetric D3-brane configuration depicted in
fig.\ref{PictureBranes}, it has been argued in \cite{CPV} that the
D3-brane stacks, positioned at $X^5=\pm l_\sigma$ in string-frame,
each lead to a half-infinite $\ads_5$ patch in the effective
5-dimensional description. Even though one starts with a
compactification on a compact 6-manifold $K_6$ for which the range
of $X^5$ is compact, the throats are governed by warp-factors
which can map a compact $X^5$ domain into a semi-infinite
non-compact $x^5$ range. Note that the D3-branes are 4-dimensional
spacetime filling and appear thus as points on the
compactification manifold.\footnote{For approaches to embed
effective brane configurations into supergravity see \cite{BC1},
\cite{BC2}, \cite{KL}, \cite{CLP}.}

Because the D3-branes appear as points on the internal 6-manifold
$K_6$ (or elliptically fibered Calabi-Yau four-fold in the
F-theory description) they are not sensitive to the global
properties of the compactification manifold. It is only the number
of D3-branes which has an influence on global properties via the
tadpole cancellation condition \cite{SVW},\cite{GVW} which
expresses the conservation of Ramond-Ramond (RR) 5-form flux. For
F-theory compactifications on an elliptically fibered Calabi-Yau
fourfold $K_8$ with $K_6$ as the base for the corresponding IIB
compactification, the tadpole cancellation condition states that
the Euler-characteristic $\chi$ of $K_8$, the background fluxes
and the number of D3-branes $N_{D3}$ have to satisfy
\begin{equation}
  N_{D3} =   \frac{\chi(K_8)}{24}
           - \int_{K_6} \frac{1}{2i\tau_2} H \wedge {\bar H} \; .
\end{equation}
In our case $N_{D3}=6$ and the 3-forms $H,{\bar H}$ are given as
linear combinations of the RR and Neveu-Schwarz (NS) 3-form
field-strengths
\begin{equation}
 H = H^{RR}-\tau H^{NS} \; , \quad
 {\bar H} = H^{RR}-{\bar \tau} H^{NS} \; ,
\end{equation}
with $\tau=\tau_1+i\tau_2=a+ie^{-\phi}$ the axio-dilaton modulus
of the elliptic fibration containing axion $a$ and dilaton $\phi$.

The open strings stretching between both D3-brane stacks will lead
to massive states with mass
\begin{equation}
M_{\sigma, \text{open}} = 2 l_\sigma T
\end{equation}
when measured in string-frame. Here $T=(2\pi\alpha')^{-1}$ is the
string-tension. We would like to know what the corresponding mass
is when it is measured in the low-energy frame where an
experimentalist would access it. For this we have to relate the
IIB string-frame metric $G^\sigma_{AB}\; ; \; A,B=1,\hdots,10$ and
the low-energy metric $G_{AB}$ which is used to measure length in
the 5-dimensional scenario. This is done via the Einstein-frame
metric \cite{CPV}. In ten dimensions string-frame and
Einstein-frame metric are related through
\begin{equation}
  G^E_{AB}=e^{-\frac{\phi}{2}}G^\sigma_{AB} \; ,
\end{equation}
whereas the low-energy metric ($m,n=6,\hdots,10$)
\begin{equation}
  ds^2 = G_{AB}dx^A dx^B
       = e^{-A(x^5)}g_{\mu\nu}(x^\rho)dx^\mu dx^\nu + (dx^5)^2
          + h_{mn}(x^5,y^k)dy^m dy^n
\end{equation}
is related to $G^E_{AB}$ by a further Weyl-rescaling involving the
internal 5-dimensional volume \cite{CPV}
\begin{equation}
  G_{AB}=V_5^{1/4} G^E_{AB} \; , \qquad
  V_5 = \frac{1}{L^5_{Pl}} \int_{K_5}d^5y\sqrt{h} \; , \qquad
  h = \det h_{mn} \; .
  \label{Volume}
\end{equation}
Here $L_{Pl}=g_s^{1/4}\sqrt{\alpha^\prime}(2\pi)^{7/8}$ denotes
the 10-dimensional Planck-length with string coupling constant
$g_s=e^\phi$ and Regge slope $\alpha'$. $K_5$ stands for those
5-dimensional sections of the base-manifold $K_6$ for which $x^5$
is held constant. The effect of these rescalings is a simple
expression for the 5-dimensional Planck-mass $M_5$ in terms of
$L_{Pl}$, which can be read off from the Einstein-Hilbert action
\begin{equation}
  -\frac{1}{(2\pi)^7(\alpha^\prime)^4}\int d^{10}X\sqrt{-G^\sigma}R^\sigma
   = -\frac{1}{L_{Pl}^8}\int d^{10}x\sqrt{-G^E}R^E
   = -\frac{1}{L_{Pl}^8}\int d^{10}x\frac{\sqrt{-G}R}{V_5}
\end{equation}
and upon dimensional reduction to five dimensions leads to the
identification
\begin{equation}
  M_5 = L_{Pl}^{-1} \; .
\end{equation}
Substituting this result for $M_5$ into equ.~(\ref{EffectiveMass})
allows us to determine the value of the 10-dimensional
Planck-length
\begin{equation}
  L_{Pl} = \Big(\frac{2(1+kl)}{k M_4^2}\Big)^{1/3}
           \simeq \Big(\frac{2l}{M_4^2}\Big)^{1/3}
           = \frac{1}{(M_4^2 M_{\text{GUT}})^{1/3}}
           = \frac{1}{4\times 10^{17}\text{GeV}} \; ,
  \label{PlanckConstraint}
\end{equation}
where we have used $2l = M_{\text{GUT}}^{-1}$ which implies $2kl =
k/M_{\text{GUT}} \gg 1$ for generic values $k$ around $M_{Pl}$ and
have identified $M_4$ with its actual value $M_4
= M_{Pl}/\sqrt{16\pi} = 1.7\times 10^{18}\,$GeV. Having determined the value for
$L_{Pl}$ (and at the same time for $M_5$), we obtain from its
expression in terms of string-theory parameters the following
relation between the string-scale $M_s = 1/\sqrt{\alpha^\prime}$
and the string-coupling constant
\begin{equation}
  M_s = 4(2\pi)^{7/8}g_s^{1/4}\times 10^{17} \,\text{GeV}
      = 2\times 10^{18} g_s^{1/4} \,\text{GeV} \; .
  \label{FirstRelation}
\end{equation}

Moreover, we find from the relation between the string-frame and
low-energy metric that the inter 3-brane distance $2l$ in the
5-dimensional description and the corresponding length $2l_\sigma$
in the string description are related by
\begin{equation}
  2l=V_5^{1/8}e^{-\frac{\phi}{4}}2l_\sigma \; .
\end{equation}
An open string stretching between both D3-brane stacks gives rise
to a massive state. In the low-energy frame this mass then becomes
\begin{equation}
  M_{\text{open}} = V_5^{1/8}e^{-\frac{\phi}{4}}2l_\sigma T
                  = 2 l T
                  = \frac{M_s^2}{2\pi M_{\text{GUT}}}\; .
   \label{StringyMass}
\end{equation}
For given $g_s$ the string-scale and the mass of the open string
states is therefore fixed. We present them in table~\ref{Masses}
for various values of $g_s$. For values of $g_s$ larger than
$10^{-6}$ the open string state masses lie at or above the GUT
scale with similar values for the string-scale.

The result that the open string state masses exceed the GUT scale
\begin{table}[t]
\begin{center}
\begin{tabular}{|c||c|c|}
\hline $g_s$ & $M_s/M_{\text{GUT}}$ & $M_{\text{open}}/M_{\text{GUT}}$ \\
\hline
\hline $10^{-1}$ & $56.2$ & $502.0$ \\
\hline $10^{-2}$ & $31.6$ & $158.7$ \\
\hline $10^{-3}$ & $17.8$ & $50.2$ \\
\hline $10^{-4}$ & $10.0$ & $15.9$ \\
\hline $10^{-5}$ & $5.6$ & $5.0$ \\
\hline $10^{-6}$ & $3.2$ & $1.6$ \\
\hline $10^{-7}$ & $1.8$ & $0.5$ \\
\hline
\end{tabular}
\caption{The string-scale $M_s$ and the mass of open string states
$M_{\text{open}}$ in units of the GUT scale
$M_{\text{GUT}}=2\times 10^{16}\,$GeV for various perturbative
values of the string coupling constant $g_s$.}
\label{Masses}
\end{center}
\end{table}
is indeed welcome because we will see that these open string
states will also give rise to the grand unified leptoquark $X$ and
$Y$ gauge bosons which mediate proton decay. Already at tree-level
the proton's lifetime, which is given by the inverse of its decay
width $\Gamma_p$, turns out to be proportional to the fourth power
of the $X$ and $Y$ mass
\begin{equation}
\Gamma_p^{-1} \propto M_{\text{open}}^4 \; .
\end{equation}
A mass for the stretched open string states and therefore the
leptoquark gauge bosons larger than the GUT scale will therefore
raise the proton's lifetime and could be crucial to avoid conflict
of supersymmetric GUT theories with proton decay experiments. More
specifically in supergravity SU(5) GUT theories there exist
dimension five baryon violating terms in the Lagrangean. They give
rise to effective dimension six operators which allow for several
decay modes of the proton. However, they are all suppressed by the
mass of the color triplet Higgs boson \cite{ACN}. We will see that
the color triplet Higgs boson originates as well from an open
string stretched between both D3-brane stacks so that its mass is
also given by $M_{\text{open}}$. Detailed studies (for a review
see \cite{Mohapatra}) imply that $M_{\text{open}} >
M_{\text{GUT}}$ by almost a factor of 10 for consistency with
present observations on the proton's lifetime. This hierarchy is
quite unpleasant in supersymmetric field theory SU(5) GUT models
as it requires that some couplings in the superpotential
unnaturally have to be much larger than one \cite{Mohapatra}. On
the other hand we readily obtain the required hierarchy between
$M_{\text{open}}$ and $M_{\text{GUT}}$ for not too small values of
$g_s$, ensuring a longer lifetime for the proton.

We also see that the string-scale comes out close to its
traditional high value. Low string-scale scenarios in which $M_s$
is lowered to the TeV scale or the intermediate scale
$10^{11}\,\text{GeV}$ require a considerable finetuning of $g_s$
to very small values and would be in conflict with observation
since $M_{\text{open}}$ would likewise be much smaller in these
scenarios leading to rapid proton decay. We will next study the
spectrum of possible matter and gauge field states arising from
the open strings in the D3-brane picture.

\section{Gauge Fields, Higgs Fields and Doublet-Triplett Splitting}

\begin{figure}[t]
              \begin{center}
               \begin{picture}(185,160)(0,-30)
                  \Line(-30,0)(-30,100)
                  \Line(-25,0)(-25,100)
                  \Line(-20,0)(-20,100)
                  \Text(-25,-10)[]{\footnotesize SU(3)}

                  \Line(160,0)(160,100)
                  \Text(160,-10)[]{\footnotesize U(1)}

                  \Line(210,0)(210,100)
                  \Line(215,0)(215,100)
                  \Text(213,-10)[]{\footnotesize SU(2)}

                  \ArrowLine(-20,75)(160,75)
                  \Text(70,85)[]{${\bf (3,1)}_{-x}$}

                  \ArrowLine(160,35)(-20,35)
                  \Text(70,45)[]{${\bf ({\bar 3},1)}_x$}

                  \ArrowLine(210,55)(160,55)
                  \Text(185,65)[]{${\bf (1,2)}_{-y}$}

                  \ArrowLine(160,15)(210,15)
                  \Text(185,25)[]{${\bf (1,2)}_y$}

               \end{picture}
               \caption{The relevant open strings of the
                        D3-brane setup. The state ${\bf (n,m)}_z$ transforms
                        as $\bf{n}$ under SU(3), as $\bf{m}$ under SU(2) and
                        bears U(1) charge $z$. For better visibility we
                        magnify the small split into $2+1$ branes of the
                        second D3-brane stack. Note that the fundamental
                        {\bf 2} of SU(2) is pseudoreal.}
               \label{PictureOpenStrings}
              \end{center}
            \end{figure}
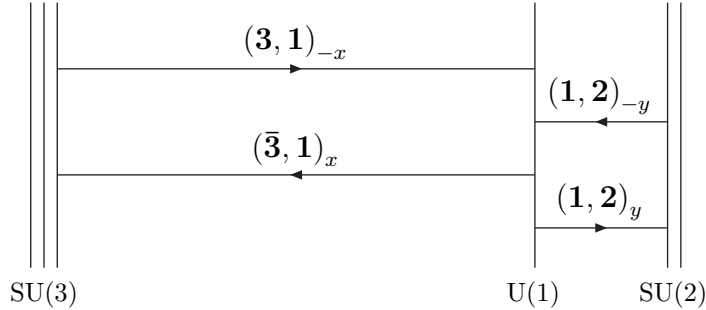

For the two D3-brane stacks with a small split of the second stack
into $2+1$ D3-branes, we will now examine the open strings, with
two orientations each, which are depicted in
fig.\ref{PictureOpenStrings} together with their transformation
properties under SU(3) and SU(2). In what follows open strings
connecting the SU(3) and the SU(2) brane stacks directly won't
play a role. They might be projected out by the imposition of an
appropriate discrete symmetry. Let us first determine the U(1)
brane charges $x$ and $y$ of the two open string states of
fig.\ref{PictureOpenStrings}. For this imagine in a
Gedankenexperiment that all six D3-branes were initially placed on
top of each other giving rise to a U(6) gauge group. The
Chan-Paton degrees of freedom at the endpoints of an open string
attached to this U(6) brane stack transform as ${\bf 6}$
resp.~${\bf\bar 6}$. Hence the open string state transforms under
the adjoint ${\bf 6}\times{\bf\bar 6}$. By separating the
D3-branes into the positions of fig.\ref{PictureOpenStrings} we
break the U(6) gauge symmetry into U(3)$\times$U(2)$\times$U(1).
To determine the U(1) charges $x$ and $y$ we have to look for
states in the product ${\bf 6}\times{\bf\bar 6}$ which transform
as $\bf(3,1)$ resp.~$\bf (1,2)$ under the SU(3)$\times$SU(2) part
of the first two factors and read off their U(1) charge (the two
U(1) factors contained in U(3) and U(2) correspond to the
center-of-mass motion of the corresponding brane stacks.
When coupled to gravity these decouple from the low-energy
spectrum as the corresponding photons acquire mass at the string
scale. The remaining low-energy gauge group
SU(3)$\times$SU(2)$\times$U(1) can be thought of as arising from a
broken SU(5) GUT which we will discuss below). This yields
\begin{equation}
x=2\; , \qquad y=3 \; .
\end{equation}
We have thus two open string states
\begin{equation}
   \zeta_u = {\bf (3,1)}_{-2} \;,\; \bar\zeta_d = {\bf ({\bar 3},1)}_2
\end{equation}
with mass at or above the GUT scale and two states
\begin{equation}
   H_u = {\bf (1,2)}_3 \;,\; \bar H_d = {\bf (1,2)}_{-3}
\end{equation}
with almost vanishing mass, say at the TeV scale or less,
controlled by the little split
            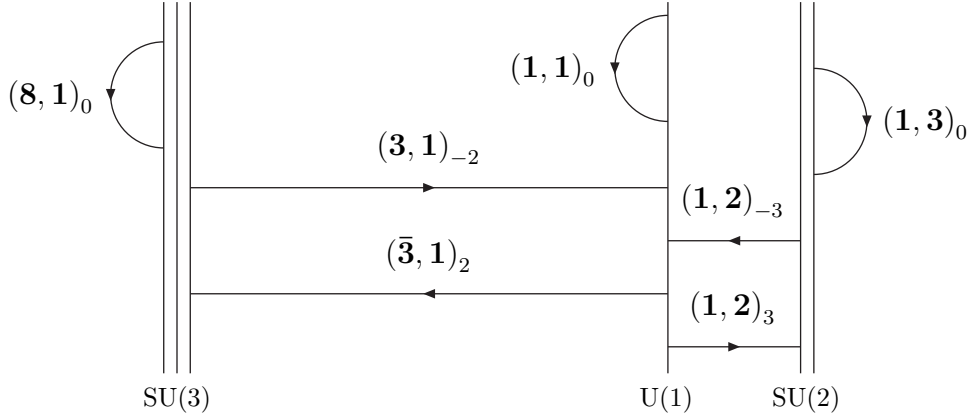
\begin{figure}[t]
              \begin{center}
               \begin{picture}(185,200)(0,-30)
                  \Line(-30,0)(-30,140)
                  \Line(-25,0)(-25,140)
                  \Line(-20,0)(-20,140)
                  \Text(-25,-10)[]{\footnotesize SU(3)}

                  \Line(160,0)(160,140)
                  \Text(160,-10)[]{\footnotesize U(1)}

                  \Line(210,0)(210,140)
                  \Line(215,0)(215,140)
                  \Text(213,-10)[]{\footnotesize SU(2)}

                  \ArrowArc(-30,105)(20,90,270)
                  \Text(-73,105)[]{${\bf(8,1)}_0$}

                  \ArrowArc(160,115)(20,90,270)
                  \Text(117,115)[]{${\bf(1,1)}_0$}

                  \ArrowArcn(215,95)(20,90,-90)
                  \Text(258,95)[]{${\bf(1,3)}_0$}

                  \ArrowLine(-20,70)(160,70)
                  \Text(70,85)[]{${\bf(3,1)}_{-2}$}

                  \ArrowLine(160,30)(-20,30)
                  \Text(70,45)[]{${\bf({\bar 3},1)}_2$}

                  \ArrowLine(210,50)(160,50)
                  \Text(185,65)[]{${\bf(1,2)}_{-3}$}

                  \ArrowLine(160,10)(210,10)
                  \Text(185,25)[]{${\bf(1,2)}_3$}

               \end{picture}
               \caption{Open strings and their SU(3)$\times$SU(2)$\times$U(1)
               transformations. Those which start and end on the same D3-brane
               stack will give rise to SM gauge fields and part of the Higgs
               bosons filling the $\bf 24$ adjoint Higgs multiplet
               of a supersymmetric SU(5) GUT. Those open strings which
               stretch between different D3-branes will provide the $\bf 5$
               and $\bf\bar 5$ SU(5) Higgs multiplets. Gauge fields
               originate from open strings vibrating in the non-compact
               directions with Neumann boundary conditions at their ends
               while Higgs fields originate from open strings vibrating in
               the internal directions satisfying Dirichlet boundary conditions on the
               D3-branes. The geometry directly implies a mass hierarchy
               between color triplet and weak doublet Higgs fields.}
               \label{SimpleStrings}
              \end{center}
            \end{figure}
between the U(1) and the SU(2) branes (see
fig.\ref{SimpleStrings}).

Depending on whether we consider open strings in the four
non-compact directions with Neumann boundary conditions along the
D3-branes or in the six compact directions with Dirichlet boundary
conditions transverse to the D3-branes, these give rise to either
gauge bosons (vector superfields) or Higgs scalars (chiral
superfields). We have denoted by $\zeta_u,\bar\zeta_d$ the color
triplet Higgs bosons and by $H_u,\bar H_d$ the weak doublet Higgs
bosons and will see later that they fit into multiplets of a
supersymmetric SU(5) GUT theory with gauge symmetry broken
spontaneously to SU(3)$\times$SU(2)$\times$U(1). As a consequence
of the large separation between the SU(3) D3-branes and the U(1)
D3-brane and on the other hand the short separation between the
SU(2) D3-branes and the U(1) D3-brane, a {\em mass-hierarchy
between color triplets states and weak doublet states} follows
directly from the geometry. We will come back to this hierarchy
later when we identify these states with SU(5) Higgs fields and
come to the doublet-triplet splitting problem.

In addition, we have also open strings which start and end on the
same stack of D3-branes. In the NS sector these give rise to
4-dimensional gauge-fields $A^{ij,\mu} =
b^\mu_{-1/2}|k_4;i,j\rangle$ ($i,j$ representing the Chan-Paton
labels) in the non-compact directions with Neumann boundary
conditions at the open string endpoints. The momenta $k_4$ are
along the four non-compact directions filled by the D3-branes.
Altogether this leads to three massless gauge-bosons

       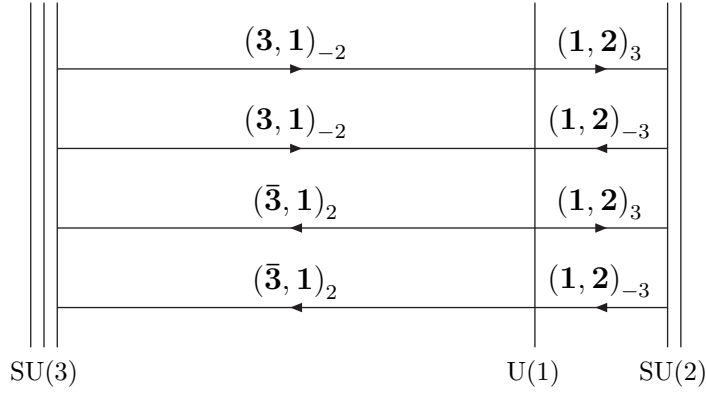
\begin{figure}[t]
        \begin{center}
         \begin{picture}(185,200)(0,-30)
                  \Line(-30,0)(-30,130)
                  \Line(-25,0)(-25,130)
                  \Line(-20,0)(-20,130)
                  \Text(-25,-10)[]{\footnotesize SU(3)}

                  \Line(160,0)(160,130)
                  \Text(160,-10)[]{\footnotesize U(1)}

                  \Line(210,0)(210,130)
                  \Line(215,0)(215,130)
                  \Text(213,-10)[]{\footnotesize SU(2)}

                  \ArrowLine(-20,105)(160,105)
                  \Text(70,115)[]{${\bf(3,1)}_{-2}$}
                  \ArrowLine(160,105)(210,105)
                  \Text(185,115)[]{${\bf(1,2)}_3$}

                  \ArrowLine(-20,75)(160,75)
                  \Text(70,85)[]{${\bf(3,1)}_{-2}$}
                  \ArrowLine(210,75)(160,75)
                  \Text(185,85)[]{${\bf(1,2)}_{-3}$}

                  \ArrowLine(160,45)(-20,45)
                  \Text(70,55)[]{${\bf({\bar 3},1)}_2$}
                  \ArrowLine(160,45)(210,45)
                  \Text(185,55)[]{${\bf(1,2)}_3$}

                  \ArrowLine(160,15)(-20,15)
                  \Text(70,25)[]{${\bf({\bar 3},1)}_2$}
                  \ArrowLine(210,15)(160,15)
                  \Text(185,25)[]{${\bf(1,2)}_{-3}$}
               \end{picture}
               \caption{States which arise from the
               composition of two stretched open strings.}
               \label{ComposedStrings}
              \end{center}
            \end{figure}

\begin{equation}
   B={\bf (1,1)}_0 \; , \;
   W^i={\bf (1,3)}_0 \; , \;
   A^\alpha={\bf (8,1)}_0
\end{equation}
which can be identified with the SM gauge bosons (see
fig.\ref{SimpleStrings}). In addition in the compact directions
transverse to the D3-branes the open strings satisfy Dirichlet
boundary conditions and give rise to scalar fields $\phi^{ij,m}
= b_{-1/2}^m|k_4;i,j\rangle\, ; \, m = 4,\hdots,9$, parameterizing the
positions of the D3-branes. These states transform in the same way
as the gauge fields under SU(3)$\times$SU(2)$\times$U(1)
\begin{equation}
\sigma_1 = {\bf (1,1)}_0 \; , \; \sigma_2 = {\bf (1,3)}_0 \; , \;
\sigma_3 = {\bf (8,1)}_0
\end{equation}
and will later build part of the SU(5) adjoint $\bf 24$ Higgs
multiplet.

Finally we can generate four more states through the composition
of two stretched open strings. These composite states, which are
marginally stable due to supersymmetry, will also have large
masses $M_{\text{open}}$ at or above the GUT scale. The four
different possibilities depicted in fig.\ref{ComposedStrings} give
us the states (similar composites appeared also recently in
\cite{KCT2}, \cite{KBH})
\begin{alignat}{3}
  {\bf (3,1)}_{-2} &\otimes {\bf (1,2)}_3 &&\rightarrow
  {\bf (3,2)}_1 \\
  {\bf (3,1)}_{-2} &\otimes {\bf (1,2)}_{-3} &&\rightarrow
  {\bf (3,2)}_{-5} \\
  {\bf (\bar{3},1)}_2 &\otimes {\bf (1,2)}_3 &&\rightarrow
  {\bf (\bar{3},2)}_5 \\
  {\bf (\bar{3},1)}_2 &\otimes {\bf (1,2)}_{-3} &&\rightarrow
  {\bf (\bar{3},2)}_{-1} \; .
\end{alignat}
Again, depending on whether we consider the strings in the
non-compact or the compact directions we will get gauge bosons or
Higgs fields with the indicated SU(3)$\times$SU(2)$\times$U(1)
transformation properties. Two of them
\begin{equation}
X = {\bf(3,2)}_{-5} \; , \; \bar Y = {\bf(\bar{3},2)}_5
\end{equation}
will account for the twelve heavy $X$ and $\bar Y$ leptoquark
gauge bosons of SU(5) and the remaining Higgs bosons
\begin{equation}
\sigma_4 = {\bf(3,2)}_{-5} \; , \; \sigma_5 = {\bf(\bar{3},2)}_5
\end{equation}
needed to fill the adjoint $\bf 24$ Higgs multiplet of SU(5).

\section{Light MSSM Matter Fields}

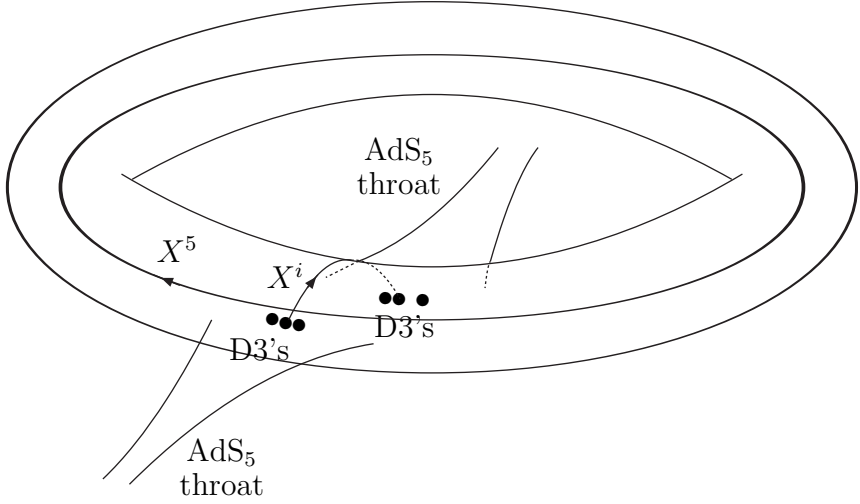
\begin{figure}[t]
\begin{center}
\begin{picture}(200,220)(0,-30)
\Oval(100,100)(70,160)(0)
\Curve{(-17,105)(100,70)(217,105)}
\Curve{(-13,103)(100,135)(213,103)}

\Text(40,50)[]{$\bullet$}
\Text(45.1,48.8)[]{$\bullet$}
\Text(50.1,47.93)[]{$\bullet$}
\Text(35,39)[]{D3's}

\Text(82.8,58.1)[]{$\bullet$}
\Text(88,57.6)[]{$\bullet$}
\Text(97,57.35)[]{$\bullet$}
\Text(90,48)[]{D3's}

\Oval(100,100)(50,140)(0)
\ArrowLine(5,63)(-5,67)
\Text(4,77)[]{$X^5$}

\Curve{(46,50)(58,68)(70,72.5)}
\DashCurve{(70,72.5)(84,64)(88,58)}{1}
\ArrowLine(53,62)(57,67)
\Text(45,67)[]{$X^i$}

\Curve{(-24,-10)(0,20)(17,50)}
\Curve{(-14,-12)(30,23)(78,41)}
\Text(21,0)[]{$\ads_5$}
\Text(21,-12)[]{throat}

\Curve{(72,72)(95,85)(125,115)}
\DashCurve{(60,66)(66,69)(72,72)}{1}
\Curve{(122,72)(127,88)(140,115)}
\DashCurve{(120,62)(121,69)(122,72)}{1}
\Text(88,114)[]{$\ads_5$}
\Text(88,102)[]{throat}
               \end{picture}
               \caption{On non-simply connected compactification
                        spaces light MSSM matter fields can arise from
                        open strings connecting D3-branes via the
                        short paths along directions $X^i$, while heavy
                        GUT fields are generated
                        from open strings connecting the same
                        D3-branes via the long path around $X^5$.
                        The D3-branes appear as points on the
                        compactification space.}
               \label{PictureSMGUT}
              \end{center}
            \end{figure}

So far we have seen how gauge and Higgs fields can arise from open
strings in the D3-brane setup. But we still need to incorporate
light matter fields (with mass at or below the TeV scale) which
could account for the matter content of the minimal supersymmetric
Standard Model (MSSM). The mechanism by which these states can
emerge along with the heavy GUT states is rather simple. Suppose
that the compactification manifold $K_6$ has non-trivial
fundamental group $\pi_1(K_6)\ne 0$. This is the case e.g.~for
orbifold compactifications which are based on torus
compactifications. Already the $n$-torus $T^n$ has a non-trivial
fundamental group
$\pi_1(T^n)=\mathbf{Z}\oplus\hdots\oplus\mathbf{Z}$ with $n$
summands $\mathbf{Z}$. The geometry will then look like in
fig.\ref{PictureSMGUT} and we can orient the direction $X^5$
around the non-simply connected path. While the open strings which
we have been discussing so far stretch around $X^5$ and therefore
have to wind around the full loop to connect both stacks of
D3-branes, we can also have very short open strings which connect
the two stacks via the orthogonal dimensions $X^6,\hdots,X^9$.
Depending
on the distance between the D3-brane stacks in the directions
$X^6,\hdots,X^9$ the mass of these ``short'' open strings can be
made very small. This represents a simple way to obtain both heavy
GUT excitations and light matter fields from open strings
stretching between the same stacks of D3-branes.

\begin{figure}[t]
              \begin{center}
               \begin{picture}(185,190)(50,-30)
                  \Line(30,0)(30,140)
                  \Text(30,-10)[]{\footnotesize U(1)}

                  \Line(88.5,0)(88.5,140)
                  \Line(91.5,0)(91.5,140)
                  \Text(90,-10)[]{\footnotesize SU(2)}

                  \Line(150,0)(150,140)
                  \Line(153,0)(153,140)
                  \Line(156,0)(156,140)
                  \Text(153,-10)[]{\footnotesize SU(3)}

                  \ArrowLine(88.5,130)(30,130)
                  \Text(170,130)[l]{$L={\bf (1,2)}_{-3}$}

                  \ArrowLine(30,100)(86,100)
                  \CArc(90,100)(4,0,180)
                  \ArrowLine(94,100)(150,100)
                  \Text(170,100)[l]{${\bar D}={\bf ({\bar 3},1)}_2$}

                  \ArrowLine(86,70)(30,70)
                  \CArc(90,70)(4,0,180)
                  \ArrowLine(150,70)(94,70)
                  \ArrowLine(30,65)(88.5,65)
                  \Text(170,68.5)[l]{$Q={\bf (3,1)}_{-2}\otimes{\bf (1,2)}_3$}

                  \ArrowLine(86,40)(30,40)
                  \CArc(90,40)(4,0,180)
                  \ArrowLine(150,40)(94,40)
                  \ArrowLine(86,35)(30,35)
                  \CArc(90,35)(4,0,180)
                  \ArrowLine(150,35)(94,35)
                  \Text(170,38.5)[l]{${\bar U}\subset{\bf (3,1)}_{-2}\otimes
                                     {\bf (3,1)}_{-2}$}

                  \ArrowLine(30,10)(88.5,10)
                  \ArrowLine(30,5)(88.5,5)
                  \Text(170,8.5)[l]{${\bar E}\subset{\bf (1,2)}_3\otimes
                                     {\bf (1,2)}_3$}
               \end{picture}
               \caption{Light MSSM quark and lepton chiral superfields
                        $L,\bar D,Q,\bar U,\bar E$ arise from open string
                        states $(L,{\bar D})$ and compositions
                        thereof $(Q,{\bar U},{\bar E})$. The open strings
                        connect the D3-branes via the short paths
                        along directions $X^i\ne X^5$.}
               \label{SMStrings}
              \end{center}
            \end{figure}
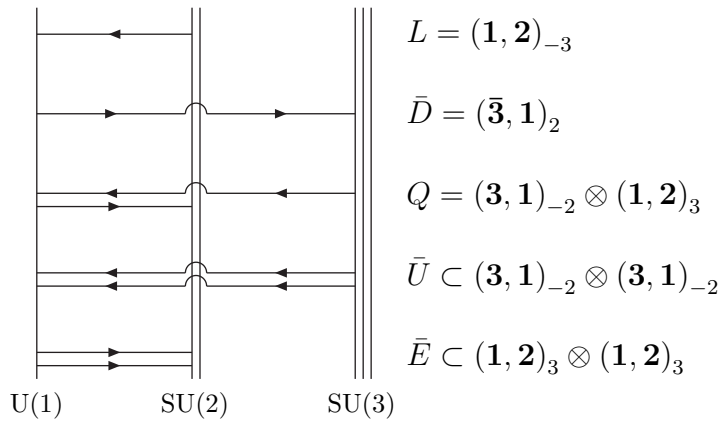

The transformation of the ``short'' open string states under
SU(3)$\times$SU(2)$\times$U(1) follows the same lines as discussed
before. In particular we can identify the MSSM lepton and quark
chiral superfields $L, \bar D, Q, \bar U, \bar E$ with the open
strings shown in fig.\ref{SMStrings}. $L$ and $\bar D$ arise
directly from two open strings
\begin{equation}
L = {\bf (1,2)}_{-3} \; , \; \bar D = {\bf (\bar 3,1)}_2
\end{equation}
while $Q, \bar U, \bar E$ arise from their compositions. More
specifically the $Q, {\bar U}, {\bar E}$ chiral superfields arise
from the following compositions
\begin{alignat}{3}
   Q &= {\bf (3,2)}_1 = {\bf (3,1)}_{-2} \otimes {\bf (1,2)}_3 \\
   {\bar U} &= {\bf ({\bar 3},1)}_{-4} \subset {\bf (3,1)}_{-2} \otimes
                                               {\bf (3,1)}_{-2} \\
   {\bar E} &= {\bf (1,1)}_6 \subset {\bf (1,2)}_3 \otimes
                                     {\bf (1,2)}_3 \; ,
\end{alignat}
whose geometrical meaning is given in fig.\ref{SMStrings}. For
this we have used ${\bf 3}\otimes {\bf 3}= {\bf {\bar 3}} + {\bf
6}$ and ${\bf 2} \otimes {\bf 2} = {\bf 1} + {\bf 3}$ in the last
two cases and picked the antisymmetric part while dismissing the
symmetric one.

This completes our discussion of the generic features of the
original 5-dimensional 3-brane configuration when lifted to IIB
string-theory with D3-branes. We have seen that heavy and light
gauge and Higgs fields arise with a mass hierarchy dictated by the
geometry. Also light MSSM matter chiral superfields with the
correct SU(3)$\times$SU(2)$\times$U(1) transformation can be
accommodated. We will next show how these states in fact combine
into multiplets of a supersymmetric GUT theory with gauge group
SU(5). Of course the states which we have discussed constitute a
subset of all in principle possible string-states. It will be left
to a case by case model analysis to apply suitable discrete
projections to get rid of unwanted additional states. This is,
however, beyond the scope of the present paper whose concern is
with general properties. There is also the important issue of
breaking the initial ${\cal N}=4$ supersymmetry preserved by the
D3-brane stacks to ${\cal N}=1$. This will be discussed in the
next but one section.

\section{SU(5) Grand Unification}

Let us now show that the string states described so far actually
fill out the multiplets of a supersymmetric SU(5) grand unified
theory. The gauge group is, however, spontaneously broken to
SU(3)$\times$SU(2)$\times$U(1). This breaking is spontaneous
because the masses are generated in the string-theory description
by separating D3-branes from each other, a mechanism which is
known to correspond to a Higgs mechanism in the effective field
theory description.

It is clear that when all six D3-branes would lie on top of each
other an unbroken SU(6) gauge group would be restored. One
therefore expects that when the branes are separated into three
SU(3) branes, two SU(2) branes and one U(1) brane, as in our case,
that one should be able to recover the spectrum of a
supersymmetric GUT with gauge group SU(6), or a subgroup thereof,
broken spontaneously to the SM gauge group. This identification
can indeed be carried out for SU(6) but is complicated by the fact
that the Higgs and matter content required for a supersymmetric
SU(6) GUT is quite involved and includes several
SU(3)$\times$SU(2)$\times$U(1) singlets \cite{GUT}. We will
therefore focus on the supersymmetric SU(5) GUT theory which has
in its minimal formulation a quite succinct field content.

The minimal field content for a supersymmetric SU(5) GUT requires,
next to a vector superfield transforming as the adjoint $V^a = \bf
24$, chiral superfields for three generations of chiral fermions
transforming as the $\psi^f = \bf \bar 5$ and $\chi^f
= \bf 10$ ($f=1,2,3$ labeling the families). In addition there is
the adjoint $\Sigma^a = \bf 24$ of scalars necessary to break the
SU(5) symmetry to the SM gauge group and two further fundamental
and anti-fundamental Higgs scalars $H=\bf 5$, $\bar H=\bf\bar 5$.
These are associated with the electroweak symmetry breaking. The
SU(5) field content can therefore be summarized as
\begin{itemize}
  \item Gauge: \hspace{1mm}$V^a = \bf 24$
  \item Matter: $\chi^f = {\bf 10}$, $\psi^f = {\bf \bar 5}$
  \item Higgs: \hspace{2mm}$\Sigma^a = \bf 24$, $H=\bf 5$, $\bar H=\bf\bar 5$
\end{itemize}

Broken down to the SM gauge group, we have the following
decompositions for the fundamental {\bf 5}, the antisymmetric
tensor {\bf 10} and adjoint {\bf 24} under
$\text{SU(3)}\times\text{SU(2)}\times\text{U(1)}$
\begin{alignat*}{3}
   &{\bf 24} = {\bf (1,1)}_0 + {\bf (1,3)}_0 + {\bf (8,1)}_0 +
               {\bf (3,2)}_{-5} + {\bf ({\bar 3},2)}_5 \\
   &{\bf 10} = {\bf (1,1)}_6 + {\bf ({\bar 3},1)}_{-4} + {\bf (3,2)}_1 \\
   &{\bf 5} = {\bf (1,2)}_3 + {\bf (3,1)}_{-2} \; .
\end{alignat*}
The SU(5) superfields hence break up into the following
$\text{SU(3)}\times\text{SU(2)}\times\text{U(1)}$ superfields
\begin{itemize}
  \item Gauge: \,$V^a = \bf 24$ \;$\rightarrow$\; massless: $B = {\bf (1,1)}_0$,
               $W^i = {\bf (1,3)}_0$, $A^\alpha = {\bf (8,1)}_0$ \\
               \phantom{Gauge: $V^a = \bf 24$ \;$\rightarrow$\; }
               \!massive:\, $X = {\bf (3,2)}_{-5}$,
                $\bar Y = {\bf ({\bar 3},2)}_5$
  \item Matter: $\chi^f = {\bf 10}$ \;$\rightarrow$\;
                light: ${\bar E} = {\bf (1,1)}_6$,
                ${\bar U} = {\bf ({\bar 3},1)}_{-4}$,
                $Q = {\bf (3,2)}_1$\\
                \phantom{Matter:\,}
                $\psi^f = {\bf \bar 5}$ \;\;\,$\rightarrow$\;
                light: \hspace{0.4mm}$L = {\bf (1,2)}_{-3}$,
                $\bar D = {\bf (\bar 3,1)}_2$
\end{itemize}

\newpage

\begin{itemize}
  \item Higgs:  \;\;$\Sigma^a = \bf 24$ \;$\rightarrow$\; massive:
                $\sigma_1 = {\bf (1,1)}_0$,
                $\sigma_2 = {\bf (1,3)}_0$,
                $\sigma_3 = {\bf (8,1)}_0$,\\
                \phantom{}\hspace{5.65cm}$\sigma_4 = {\bf (3,2)}_{-5}$,
                $\sigma_5 = {\bf ({\bar 3},2)}_5$ \\
                \phantom{Higgs: \;\,}
                $H = \bf 5$ \;\;\,\,$\rightarrow$\; light:
                \hspace{0.45cm}$H_u = {\bf (1,2)}_3$ \\
                \phantom{Higgs: \;\, $H = \bf 5$
                \;\;\;$\rightarrow$}\hspace{0.8mm}
                massive: \,$\zeta_u = {\bf (3,1)}_{-2}$ \\
                \phantom{Higgs: \;\,}
                $\bar H = \bf\bar 5$ \;\;\;$\rightarrow$\; light:
                \hspace{0.5cm}$\bar H_d = {\bf (1,2)}_{-3}$ \\
                \phantom{Higgs: \;\, $H = \bf 5$ \;\;\;$\rightarrow$\,}
                massive: \,$\bar\zeta_d = {\bf (\bar 3,1)}_2$
\end{itemize}
where ``massive'' means masses at or above the GUT scale while
``light'' indicates hierarchically smaller masses at or below the
TeV scale. $B,W^i,A^\alpha$ denote the vector superfields
containing the respective SM gauge bosons, while $X$ and $\bar Y$
stand for the leptoquark gauge bosons. Quarks/squarks are
contained in $Q, \bar U, \bar D$ and leptons/sleptons in $L, \bar
E$. Finally, the heavy Higgs/higgsinos come as color triplets
$\zeta_u,\bar\zeta_d$ while the light Higgs/higgsinos transform as
weak doublets, denoted by $H_u,\bar H_d$.

As we saw in detail earlier, these were precisely the states which
were generated by the open strings in the D3-brane setup.
Moreover, the states required to be ``massive'' came out indeed
with a large mass $M_{\text{open}}$ at or greater than the GUT
scale, the states labeled as ``light'' came out with small masses
and the states labeled as ``massless'' came out to be massless.
The only exception to this correct assignment of masses in the
string description are the Higgs bosons $\sigma_1, \sigma_2,
\sigma_3$ which appeared to be massless. It is at this point where
moduli stabilization would set in. Namely, the fields $\sigma_1,
\sigma_2, \sigma_3$, coming from the zero modes of the open string
components along the compact directions, represent the moduli
which describe the positions of the D3-branes. Once these
positions get stabilized, a potential will fix the values of
$\sigma_1, \sigma_2, \sigma_3$ and render them massive. These
masses will naturally take values around the string-scale which is
also at or above the GUT scale (see table~\ref{Masses}). We also
see that the U(1) can indeed be identified with the SM
$\text{U(1)}_Y$ hypercharge, and therefore remains anomaly-free,
as the U(1) charges of all fields exactly match those of the SM
$\text{U(1)}_Y$. Moreover, with the identification of the color
triplet and weak doublet states as Higgs chiral superfields, we
obtain that the 3-brane separation $2l = M_{\text{GUT}}^{-1}$,
required for a sufficient suppression of the 4-dimensional vacuum
energy, implies a simple resolution of the doublet-triplet
splitting problem in the string-theory description of the SU(5)
GUT theory. The number of families will however be model dependent
and follows from topological data of the compactification manifold
such as the Euler character.

\section{Supersymmetry Breaking}

The ${\cal N}=1$ vector and chiral superfields which made up the
gauge bosons, matter fermions and Higgs fields of the
spontaneously broken SU(5) supersymmetric GUT resp.~MSSM, came
from open strings attached to parallel D3-branes. They are thus
originating from ${\cal N}=4$ vector supermultiplets. One ${\cal
N}=4$ vector supermultiplet contains three ${\cal N}=1$ chiral
superfields plus one ${\cal N}=1$ vector supermultiplet. Let us
finally briefly address how one might break the ${\cal N}=4$
supersymmetry to an ${\cal N}=1$ supersymmetry.

The fermions originate from the Ramond-sector of the open strings
and in uncompactified ten dimensions would be described by a
Majorana-Weyl spinor $u_\alpha^{ij}|\alpha;k_{10};i,j\rangle$.
Here, $\alpha=1,\hdots,8$ is a spinor-index running over the
physical on-shell degrees of freedom, $i,j$ are the Chan-Paton
labels and $k_{10}$ is the 10-dimensional momentum. By dimensional
reduction to four dimensions, $u_\alpha$ turns into four
two-component Weyl-spinors $\lambda_a^{ij}\, ; \, a=1,\hdots,4$.
In a 4-dimensional ${\cal N}=1$ description $\lambda_4^{ij}$ gets
combined with the gauge-field $A_\mu^{ij}$ of the Neveu-Schwarz
sector into an ${\cal N}=1$ vector-superfield, while the remaining
three spinors $\lambda_1^{ij},\lambda_2^{ij},\lambda_3^{ij}$ are
each paired with two real Neveu-Schwarz scalars
$(\phi_4^{ij},\phi_7^{ij}), (\phi_5^{ij},\phi_8^{ij}),
(\phi_6^{ij},\phi_9^{ij})$ to build the three chiral superfields
\begin{equation}
Z_a^{ij} = (\phi_{a+3}^{ij}+i\phi_{a+6}^{ij},\lambda_a^{ij})\;
,\quad a
= 1,2,3 \; .
\end{equation}
Together the three chiral superfields plus the single vector
superfield make up the ${\cal N}=4$ vector supermultiplet. Hence
we naturally arrive at a multiplicity of three for the chiral
matter fermions. It would be interesting to explore in concrete
models the connection between this multiplicity and the number
three of fermion families which we will leave to future work.

One way of breaking the ${\cal N}=4$ supersymmetry to an ${\cal
N}=1$ supersymmetry is, at the field-theory level, by adding
masses to the three chiral superfields in an ${\cal N}=1$
supersymmetry preserving way \cite{Mass1}, \cite{Mass2}. For this
one supplements the ${\cal N}=1$ superpotential by the mass terms
\begin{equation}
\Delta W \sim \sum_{a=1}^3 m_a\text{tr}Z_a^2 \; .
\end{equation}
This lifts the mass degeneracy of the chiral superfields by
construction. What makes these mass perturbations not ad hoc and
even natural in string-theory, is that the fact that they
correspond to magnetic 3-form fluxes $H$ (see \cite{Mass2} for
details). As we have seen earlier such fluxes are required in
generic string-theory compactifications by the tadpole
cancellation condition.

Let us finally point out yet another way to break, with the help
of additional D7-branes, ${\cal N}=4$ to ${\cal N}=1$
supersymmetry. The generic tangent space group of a 6-dimensional
compactification manifold is SO(6). The D3-branes were all
transverse to the compactification manifold and will therefore not
influence its tangent space group. This will change, however, when
additional D7-branes are present in the type IIB string-theory
vacuum. Consider for instance three D7-branes with worldvolume
along the directions 01234578, 01234679 and 01235689. Together,
they preserve $1/8$ of the initial 32 IIB supercharges, and
therefore leave us with the desired ${\cal N}=1$ supersymmetry in
four dimensions. Moreover, the supersymmetry preserved by the
D7-branes is compatible with the one preserved by the D3-branes.
To see this, let us write down the conditions imposed on the
supersymmetry parameters $\epsilon_L,\epsilon_R$ (16-component
Majorana-Weyl spinors in IIB of the same chirality) by the
presence of the D7-branes. These are
\begin{alignat}{3}
  \epsilon_L &= \Gam_{D3}\Gam_{R_1}\epsilon_R \; ; \quad
                \Gam_{R_1}=\Gam^4\Gam^5\Gam^7\Gam^8\; , \;\;
                \Gam_{D3}=\Gam^0\Gam^1\Gam^2\Gam^3 \\
  \epsilon_L &= \Gam_{D3}\Gam_{R_2}\epsilon_R \; ; \quad
                \Gam_{R_2}=\Gam^4\Gam^6\Gam^7\Gam^9 \\
  \epsilon_L &= \Gam_{D3}\Gam_{R_3}\epsilon_R \; ; \quad
                \Gam_{R_3}=\Gam^5\Gam^6\Gam^8\Gam^9 \; .
\end{alignat}
Taken together, they imply that
\begin{equation}
  \epsilon_L = \Gam_{D3}\epsilon_R \; ,
\end{equation}
which is exactly the condition imposed by a 4-dimensional
spacetime filling D3-brane. Therefore the D3-branes preserve the
${\cal N}=1$ supersymmetry preserved by the D7-branes and can be
placed at the common 4-dimensional intersection of the three
intersecting D7-branes. The presence of the D7-branes will break
the SO(6) tangent space group down to
\begin{equation}
   \text{SO(6)} \supset \text{SO(2)}\times \text{SO(2)}\times \text{SO(2)}
                = \text{U(1)}\times \text{U(1)}\times \text{U(1)} \; .
\end{equation}
This entails a corresponding split of the three chiral
superfields, initially combined into the ${\cal N}=4$ vector
multiplet, into three separate multiplets as each one transforms
under a separate U(1). Thus the degeneracy between them gets
lifted and they can acquire different ${\cal N}=1$ masses.

\bigskip
\noindent {\large \bf Acknowledgements}\\[2ex]
We would like to thank Dieter L\"ust, Lisa Randall and Raman
Sundrum for discussion and correspondence.

\newpage

\begin{appendix}

\section{Warped Geometry and Effective D=4 Action for Unequal Wall
         Tensions}

\begin{figure}[b]
              \begin{center}
               \begin{picture}(150,140)(-10,0)
                  \LongArrow(50,9)(50,90)
                  \Text(50,104)[]{$A(x^5)$}

                  \LongArrow(-20,10)(130,10)
                  \Text(142,13)[]{$x^5$}

                  \Text(0,3)[]{$-l$}
                  \Line(0,9)(0,11)

                  \Text(50,3)[]{$0$}

                  \Text(100,3)[]{$l$}
                  \Line(100,9)(100,11)

                  \Line(-10,100)(0,37)
                  \Line(0,37)(100,41.5)
                  \Line(100,41.5)(110,100)

                  \Text(38,48)[]{$k_1 l$}
                  \Line(49,43)(51,43)

                  \Text(62,34)[]{$k_2 l$}
                  \Line(49,37)(51,37)

               \end{picture}
               \caption{The function $A(x^5)$ which determines the
                        warp-factor.}
               \label{PictureWarp}
              \end{center}
            \end{figure}
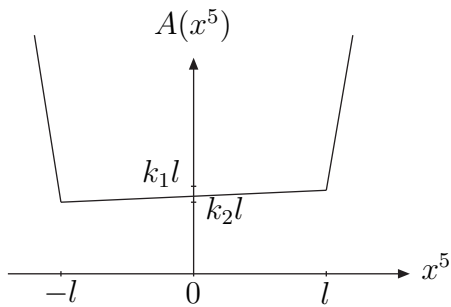

In this appendix we analyze the warped geometry and effective
action for unequal 3-brane tensions $T_1 \neq T_2$. Let us
emphasize that in string-theory all D3-branes come with the same
tension, so that equal tensions simply required an equal number of
D3-branes. The unequal tension case can nevertheless be relevant
in the effective 5-dimensional description. In this case the
Ansatz (\ref{Ansatz}) yields the solution
\begin{equation}
   A(x^5) = \frac{k_1}{2}\left|x^5+l\right|
           +\frac{k_2}{2}\left|x^5-l\right|
          = \left\{ \begin{array}{cl}
      \frac{1}{2}K_{12} x^5 +\frac{1}{2}k_{12}l
      &,\,\, x^5 \ge l \\
      \frac{1}{2}k_{12}x^5 + \frac{1}{2}K_{12}l
      &,\,\, -l \le x^5 \le l\\
      -\frac{1}{2}K_{12}x^5 -\frac{1}{2}k_{12}l
      &,\,\, x^5 \le -l \end{array}
            \right.
       \; ,
\end{equation}
with constants $K_{12}=k_1+k_2$ and $k_{12}=k_1-k_2$. Without loss
of generality, we can assume that $k_1 \ge k_2$ subsequently. The
function $A(x^5)$ which determines the warp-factor is displayed in
fig.\ref{PictureWarp}.
The corresponding warp-factor $e^{-A(x^5)}$ is bounded from above
by $e^{-k_2 l}$ over the whole fifth dimension. Again we have set
an arbitrary integration constant which could be added to $A(x^5)$
to zero. The Einstein equations (\ref{EinsteinEquation}) determine
the stepwise constant bulk cosmological constant
\begin{equation}
   \Lambda(x^5) =
               \left\{ \begin{array}{cc}
             \Lambda_e \; , & |x^5| \ge l\\
             \Lambda_i \; , & |x^5| < l
                       \end{array}
               \right.
                 = -\frac{3M_5^3}{4}
               \left\{ \begin{array}{cc}
             K_{12}^2 \; , & |x^5| \ge l\\
             k_{12}^2 \; , & |x^5| < l
                       \end{array}
               \right.
        \label{Lambda}
\end{equation}
and 3-brane tensions
\begin{equation}
T_1 = 3M_5^3 k_1 \; , \qquad T_2 = 3M_5^3k_2 \; .
        \label{Tension}
\end{equation}
The next task is again the determination of the effective
4-dimensional action by integration over the fifth dimension.
Along the same lines as before, by employing (\ref{EHReduction}),
we get for the Einstein-Hilbert term
\begin{alignat}{3}
   S_{EH} &= -\int d^4 x \sqrt{g} M_5^3 \left( R(g) \int_{-\infty}^\infty dx^5
             e^{-A} + \int_{-\infty}^\infty dx^5 e^{-2A}
             \left[ 5(A^\prime)^2-4A^{\prime\prime} \right]
                                      \right) \notag \\
          &= -e^{-K_{12}l/2} \int d^4x \sqrt{g}M_5^3
            \Bigg( 4R(g) \Big[ \frac{1}{K_{12}}\cosh\left(\frac{
                                 k_{12}l}{2}\right)
                               +\frac{1}{k_{12}}\sinh\left(\frac{
                                 k_{12}l}{2}\right)
                          \Big] \notag \\
          &\phantom{=\;}
             + \frac{5}{4} e^{-K_{12}l/2}
                           \Big[ 2 K_{12}\cosh\left(k_{12}l\right)
                                +2 k_{12}\sinh\left(k_{12}l\right)
                           \Big]  \notag \\
          &\phantom{=\;}
             - 4 e^{-K_{12}l/2}
                           \Big[ k_1 e^{k_{12}l}
                                +k_2 e^{-k_{12}l}
                           \Big]
             \Bigg) \; .
\end{alignat}
For the brane-terms and bulk cosmological constant term we get
\begin{alignat}{3}
   S_{SM_1}+S_{SM_2}+S_\Lambda = &-e^{-K_{12}l} \int d^4x \sqrt{g}
         \bigg( e^{k_{12}l}T_1 + e^{-k_{12}l}T_2 \notag \\
                               &+2\frac{\Lambda_e}{K_{12}}\cosh(k_{12}l)
                                +2\frac{\Lambda_i}{k_{12}}\sinh(k_{12}l)
                         \bigg) \; .
\end{alignat}
Pulling out an overall constant factor of $e^{-K_{12}l/2}$ in
front, the final effective action reads
\begin{alignat}{3}
   &\phantom{=\;\;}S_{EH}+S_{SM_1}+S_{SM_2}+S_\Lambda \notag \\
          &= -e^{-K_{12}l/2}\int d^4x\sqrt{g}
   \Bigg( 4M_5^3 R(g)\Big[\frac{1}{K_{12}}\cosh\left(\frac{k_{12}l}{2}\right)
                       + \frac{1}{k_{12}}\sinh\left(\frac{k_{12}l}{2}\right)
                    \Big] \notag \\
          &\phantom{=\;}
             + \frac{5}{2}M_5^3 e^{-K_{12}l/2}
                           \Big[ K_{12}\cosh\left(k_{12}l\right)
                                +k_{12}\sinh\left(k_{12}l\right)
                           \Big]
        + e^{-K_{12}l/2} \Big[ e^{k_{12}l}(T_1-4k_1 M_5^3) \notag \\
          &\phantom{=\;}
                               +e^{-k_{12}l}(T_2-4k_2 M_5^3)
                               +2\frac{\Lambda_e}{K_{12}}\cosh(k_{12}l)
                               +2\frac{\Lambda_i}{k_{12}}\sinh(k_{12}l)
                          \Big]
   \Bigg) \; .
\end{alignat}
By the same reasoning as explained in the main text, we drop the
overall constant factor and arrive at the effective action
\begin{equation}
       S_{D=4}  = -\int d^4x\sqrt{g}
   \left( M_4^2 R(g) + \Lambda_4 \right) \; ,
             \label{FourActionLong}
\end{equation}
with effective 4-dimensional Planck-scale $M_4$ and 4-dimensional
vacuum energy $\Lambda_4$ given by
\begin{alignat}{3}
   M_4^2 &= 4M_5^3
          \Big[\frac{1}{K_{12}}\cosh\Big(\frac{k_{12}l}{2}\Big)
              +\frac{1}{k_{12}}\sinh\Big(\frac{k_{12}l}{2}\Big)
          \Big]               \\
   \Lambda_4 &= e^{-K_{12}l/2}
            \bigg(
               \frac{5}{2}M_5^3
                           \Big[ K_{12}\cosh\left(k_{12}l\right)
                                +k_{12}\sinh\left(k_{12}l\right)
                           \Big]
             + \Big[ e^{k_{12}l}(T_1-4k_1 M_5^3)  \notag \\
          &\phantom{=\;\qquad\quad\;}
                    +e^{-k_{12}l}(T_2-4k_2 M_5^3)
                    +2\frac{\Lambda_e}{K_{12}}\cosh(k_{12}l)
                    +2\frac{\Lambda_i}{k_{12}}\sinh(k_{12}l)
               \Big]
            \bigg)
       \label{EffectiveCoCo2}  \; .
\end{alignat}
Again, there exists an exponential suppression-factor
$e^{-K_{12}l/2}$ multiplying the vacuum energy, which allows to
suppress $\Lambda_4$ down to its observational value for generic
values $k_1,k_2 \approx M_{Pl}$.  When the values
(\ref{Lambda}),(\ref{Tension}) for $T_1,T_2,\Lambda_e,\Lambda_i$
are substituted into the obtained effective action, we arrive at a
vanishing $\Lambda_4$, which checks the derivation of the action,
since in this case the fine-tuning of the parameters should
guarantee a flat 4-dimensional metric $g_{\mu\nu}=\eta_{\mu\nu}$.
For the particular case of coinciding brane-tensions, $T_1=T_2=T$
(which entails $k_1=k_2=k$), we arrive at the effective action
given by
(\ref{FourAction}),(\ref{EffectiveMass}),(\ref{EffectiveCoCo}),
which was discussed in the main text.

Again, let us now relax the fine-tuning
\begin{equation}
            \Lambda(x^5)
            =
               \left\{ \begin{array}{cc}
                       \Lambda_e \; , & |x^5| \ge l\\
                       \Lambda_i \; , & |x^5| < l
                       \end{array}
               \right.
            = -\frac{1}{12M_5^3}
               \left\{ \begin{array}{cc}
             (T_1+T_2)^2 \; , & |x^5| \ge l\\
             (T_1-T_2)^2 \; , & |x^5| < l
                       \end{array}
               \right. \; ,
\end{equation}
between the bulk cosmological constant and the 3-brane tensions
which implies a non-flat 4-dimensional metric $g_{\mu\nu}\neq
\eta_{\mu\nu}$ in the Ansatz
\begin{equation}
  ds^2 = e^{-A(x^5)}g_{\mu\nu}dx^\mu dx^\nu + (dx^5)^2 \; .
\end{equation}
Depending on whether $\Lambda_4$ is positive or negative, the
metric $g_{\mu\nu}$ describes either a de Sitter or anti-De Sitter
spacetime. Our interest lies in the de Sitter case. From
(\ref{EffectiveCoCo2}) it is evident, that in order to arrive at
an exponentially small $\Lambda_4$, the difference between both
3-brane tensions cannot be too large but has to be constrained by
\begin{equation}
  k_1-k_2\equiv k_{12} \lesssim \frac{1}{l}=2 M_{\text{GUT}} \; .
\end{equation}
Most naturally we would expect values such as $k_1,k_2 \approx
M_{Pl}$, $T_1,T_2 \approx M_{Pl}^4$, $\Lambda_e\approx M_{Pl}^5$
and for the 5-dimensional Planck-scale $M_5\approx M_{Pl}$. If
also $\Lambda_i$ does not exceed $(3\times 10^{18}\text{GeV})^5$,
which is a bit larger than the reduced Planck mass, then we
recognize from (\ref{EffectiveCoCo2}) that the suppression by the
exponential pre-factor is just sufficient, in view of
(\ref{PlanckGUTRelation2}), to decrease the Planck-valued
contributions to the 4-dimensional vacuum energy down to its
observed value.

\section{The Effective Potential for Bulk Scalars in the Case of Unequal
         Wall-Tensions}

For completeness, let us also derive the bulk scalar contribution
to the effective potential $\Lambda_4$ in the case with unequal
brane tensions. With the same action for the bulk scalar $\Phi$
with mass $m$ as in the main text, we obtain for unequal brane
tensions the following solution to the field equation
\begin{equation}
   \Phi(x^5)= \left\{
                   \begin{array}{cl}
                      a e^{(1+\Gamma)A(x^5)} + b e^{(1-\Gamma)A(x^5)}
                     &,\,\, x^5  < -l \;  \\
                      c e^{(1+\gamma)A(x^5)} + d e^{(1-\gamma)A(x^5)}
                     &,\,\,|x^5| \le l \\
                      e e^{(1+\Gamma)A(x^5)} + f e^{(1-\Gamma)A(x^5)}
                     &,\,\, x^5 > l
                   \end{array}
              \right. \; ,
\end{equation}
where now
\begin{equation}
  \Gamma = \sqrt{1+4m^2/K_{12}^2} \; , \qquad \gamma = \sqrt{1+4m^2/k_{12}^2}
  \; .
\end{equation}
In order to obtain a normalizable solution for $\Phi$, we set the
coefficients $a=e=0$. Moreover, imposing continuity of $\Phi$ at
the position of the 3-branes determines $b$ and $f$ in terms of
$c$ and $d$
\begin{alignat}{3}
  b&= e^{\Gamma k_2 l} \btil \; , \qquad
  \btil = c e^{\gamma k_2 l}+d e^{-\gamma k_2l}    \\
  f&= e^{\Gamma k_1 l} \ftil \; , \qquad
  \ftil = c e^{\gamma k_1 l}+d e^{-\gamma k_1l} \; .
\end{alignat}
To fix the remaining coefficients $c$ and $d$ one would have to
plug the above bulk solution into the field equation and integrate
out the fifth dimension to incorporate the brane boundary
conditions. Since this leads to a complicated cubic equation in
the unknowns $c$ and $d$, it is again easier to determine them by
inserting the bulk solution into the scalar action and integrating
it over $x^5$ to arrive at an effective potential for the
inter-brane distance $2l$. For positive couplings
$\lambda_1,\lambda_2$ this effective potential will be positive
definite. Hence, to minimize the potential, we are led to set
$\Phi(-l)=v_1$ and $\Phi(l)=v_2$. This allows for a determination
of $c$ and $d$ in terms of the vacuum expectation values $v_1,v_2$
\begin{equation}
   c=\frac{v_2 e^{-(1-\gamma)k_1l} - v_1 e^{-(1-\gamma)k_2l}}
          {e^{2\gamma k_1l} - e^{2\gamma k_2l}} \; , \qquad
   d=\frac{v_2 e^{-(1+\gamma)k_1l} - v_1 e^{-(1+\gamma)k_2l}}
          {e^{-2\gamma k_1l} - e^{-2\gamma k_2l}} \; .
\end{equation}
The effective potential\footnote{We use the relations
         $(1\pm\gamma)^2\frac{k_{12}^2}{4}+m^2=\gamma(\gamma\pm
         1)\frac{k_{12}^2}{2}$ and $(1-\gamma^2)\frac{k_{12}^2}{4}+m^2=0$.}
then eventually becomes
\begin{alignat}{3}
   V_\Phi(l) = \, &\frac{k_{12}}{2}\sinh(\gamma k_{12} l)
                \bigg( c^2(\gamma+1) e^{\gamma K_{12} l}
                      +d^2(\gamma-1) e^{-\gamma K_{12} l}
                \bigg)   \notag \\
             &+\frac{(\Gamma-1)K_{12}}{4}\big( \btil^2+\ftil^2 \big) \; .
\end{alignat}
A numerical investigation of this potential shows that, also in
the case with differing tensions, a bulk scalar leads generically
to an effective potential, which is likewise sufficiently
exponentially suppressed.
\end{appendix}

 \newcommand{\zpc}[3]{{\sl Z. Phys.} {\bf C#1} (#2) #3}
 \newcommand{\zp}[3]{{\sl Z. Phys.} {\bf #1} (#2) #3}
 \newcommand{\npb}[3]{{\sl Nucl. Phys.} {\bf B#1} (#2)~#3}
 \newcommand{\plb}[3]{{\sl Phys. Lett.} {\bf B#1} (#2) #3}
 \newcommand{\prd}[3]{{\sl Phys. Rev.} {\bf D#1} (#2) #3}
 \newcommand{\prl}[3]{{\sl Phys. Rev. Lett.} {\bf #1} (#2) #3}
 \newcommand{\prep}[3]{{\sl Phys. Rep.} {\bf #1} (#2) #3}
 \newcommand{\fp}[3]{{\sl Fortsch. Phys.} {\bf #1} (#2) #3}
 \newcommand{\nc}[3]{{\sl Nuovo Cimento} {\bf #1} (#2) #3}
 \newcommand{\ijmpa}[3]{{\sl Int. J. Mod. Phys.} {\bf A#1} (#2) #3}
 \newcommand{\rmp}[3]{{\sl Rev. Mod. Phys.} {\bf #1} (#2) #3}
 \newcommand{\ptp}[3]{{\sl Prog. Theor. Phys.} {\bf #1} (#2) #3}
 \newcommand{\sjnp}[3]{{\sl Sov. J. Nucl. Phys.} {\bf #1} (#2) #3}
 \newcommand{\splir}[3]{{\sl Sov. Phys. Leb. Inst. Rep.} {\bf #1} (#2) #3}
 \newcommand{\cpc}[3]{{\sl Comp. Phys. Commun.} {\bf #1} (#2) #3}
 \newcommand{\mpla}[3]{{\sl Mod. Phys. Lett.} {\bf A#1} (#2) #3}
 \newcommand{\cmp}[3]{{\sl Commun. Math. Phys.} {\bf #1} (#2) #3}
 \newcommand{\cqg}[3]{{\sl Class. Quant. Grav.} {\bf #1} (#2) #3}
 \newcommand{\jmp}[3]{{\sl J. Math. Phys.} {\bf #1} (#2) #3}
 \newcommand{\nim}[3]{{\sl Nucl. Instr. Meth.} {\bf #1} (#2) #3}
 \newcommand{\el}[3]{{\sl Europhysics Letters} {\bf #1} (#2) #3}
 \newcommand{\ap}[3]{{\sl Ann. of Phys.} {\bf #1} (#2) #3}
 \newcommand{\sci}[3]{{\sl Science} {\bf #1} (#2) #3}
 \newcommand{\jhep}[3]{{\sl JHEP} {\bf #1} (#2) #3}
 \newcommand{\jetp}[3]{{\sl JETP} {\bf #1} (#2) #3}
 \newcommand{\jetpl}[3]{{\sl JETP Lett.} {\bf #1} (#2) #3}
 \newcommand{\acpp}[3]{{\sl Acta Physica Polonica} {\bf #1} (#2) #3}
 \newcommand{\vj}[4]{{\sl #1~}{\bf #2} (#3) #4}
 \newcommand{\ej}[3]{{\bf #1} (#2) #3}
 \newcommand{\vjs}[2]{{\sl #1~}{\bf #2}}
 \newcommand{\hep}[1]{{\sl hep--ph/}{#1}}
 \newcommand{\desy}[1]{{\sl DESY-Report~}{#1}}


\begin{thebibliography}{99}
  \bibitem{Bahcall} N.A.~Bahcall, J.P.~Ostriker, S.~Perlmutter and
                    P.J.~Steinhardt,
                    \sci{284}{1999}{1481}, astro-ph/9906463;
  \bibitem{Klapdor} H.V.~Klapdor-Kleingrothaus, H.~Pas and A.Y.~Smirnov,
                    \prd{63}{2001}{073005}, hep-ph/0003219;
  \bibitem{WittenCosm} E.~Witten,
                       \mpla{10}{1995}{2153}, hep-th/9506101;
  \bibitem{CCT} G.T.~Horowitz and D.L.~Welch,
                \prl{71}{1993}{328}, hep-th/9302126;
                E.~Alvarez, L.~Alvarez-Gaume, J.~Barbon and Y.~Lozano,
                \npb{415}{1994}{71}, hep-th/9309039;
                A.~Krause,
                \mpla{18}{2003}{2571}, hep-th/9911197;
  \bibitem{Banks} T.~Banks,
                  hep-th/0007146;
                  E.~Witten,
                  hep-th/0106109;
                  T.~Banks,
                  \ijmpa{16}{2001}{910};
  \bibitem{KCT} A.~Krause,
                \ijmpa{20}{2005}{4055}, hep-th/0201260;
                \ijmpa{20}{2005}{2813}, hep-th/0205310;
  \bibitem{KakushadzeTye} Z.~Kakushadze and S.H.H.~Tye,
                          \npb{548}{1999}{180}, hep-th/9809147;
  \bibitem{Antoniadis} N.~Arkani-Hamed, S.~Dimopoulos and G.~Dvali,
                       \plb{429}{1998}{263}, hep-ph/9803315;
                       I.~Antoniadis, N.~Arkani-Hamed, S.~Dimopoulos and
                       G.~Dvali,
                       \plb{436}{1998}{257}, hep-ph/9804398;
  \bibitem{RS1} L.~Randall and R.~Sundrum,
                \prl{83}{1999}{3370}, hep-ph/9905221;
  \bibitem{GoldWise} W.D.~Goldberger and M.~B.~Wise,
                     \prl{83}{1999}{4922}, hep-ph/9907447;
  \bibitem{ADKS} N.~Arkani-Hamed, S.~Dimopoulos, N.~Kaloper and R.~Sundrum,
                 \plb{480}{2000}{193}, hep-th/0001197;
  \bibitem{KSS} S.~Kachru, M.~Schulz and E.~Silverstein,
                \prd{62}{2000}{045021}, hep-th/0001206;
  \bibitem{FLLN1} S.~F\"orste, Z.~Lalak, S.~Lavignac and H.P.~Nilles,
                  \plb{481}{2000}{360}, hep-th/0002164;
  \bibitem{Weinberg1} S.~Weinberg,
                      \rmp{61}{1989}{1};
  \bibitem{Weinberg2} S.~Weinberg,
                      {\it The Cosmological Constant Problems},
                      astro-ph/0005265;
  \bibitem{Carroll} S.M.~Carroll,
                    {{\sl Living Rev.~Rel.} {\bf 4} (2001) 1}, astro-ph/0004075;
  \bibitem{WittenRev} E.~Witten,
                      {\it The Cosmological Constant from the Viewpoint of
                       String Theory},
                      hep-ph/0002297;
  \bibitem{Steinhardt} P.J.~Steinhardt,
                       \plb{462}{41}{1999}, hep-th/9907080;
  \bibitem{BMQ} C.P.~Burgess, R.C.~Myers and F.~Quevedo,
                \plb{495}{2000}{384}, hep-th/9911164;
  \bibitem{Schmidhuber1} C.~Schmidhuber,
                         \npb{580}{2000}{140}, hep-th/9912156;
  \bibitem{Alwis1} S.P.~de Alwis,
                   \npb{597}{2001}{263}, hep-th/0002174;
  \bibitem{Alwis2} S.P.~de Alwis, A.T.~Flournoy and N.~Irges,
                   \jhep{0101}{2001}{027}, hep-th/0004125;
  \bibitem{PolBusso} R.~Bousso and J.~Polchinski,
                     \jhep{0006}{2000}{006}, hep-th/0004134;
  \bibitem{KakuCoCo} Z.~Kakushadze,
                     \npb{589}{2000}{75}, hep-th/0005217;
  \bibitem{Schmidhuber2} C.~Schmidhuber,
                         \npb{585}{2000}{385}, hep-th/0005248;
  \bibitem{FMRSW} J.L.~Feng, J.~March-Russell, S.~Sethi and F.~Wilczek,
                  \npb{602}{2001}{307}, hep-th/0005276;
  \bibitem{TyeWasserman} S.H.~Tye and I.~Wasserman,
                         \prl{86}{2001}{1682}, hep-th/0006068;
  \bibitem{AK4} A.~Krause,
                \jhep{0309}{2003}{016}, hep-th/0007233;
  \bibitem{FLLN2} S.~F\"orste, Z.~Lalak, S.~Lavignac and H.P.~Nilles,
                  \jhep{0009}{2000}{034}, hep-th/0006139;
  \bibitem{Kaku2} Z.~Kakushadze,
                  \plb{489}{2000}{034}, hep-th/0006215;
  \bibitem{Tasinato} H.P.~Nilles, A.~Papazoglou and G.~Tasinato,
                     \npb{677}{2004}{405}, hep-th/0309042;
  \bibitem{Erdem} R.~Erdem,
                  \plb{621}{2005}{11}, hep-th/0410063;
  \bibitem{Kwon} E.K.~Park and P.S.~Kwon,
                 hep-th/0507094;
  \bibitem{Rinaldi} O.~Corradini and M.~Rinaldi,
                    {{\sl JCAP} {\bf 0601} (2006) 20}, hep-th/0509200;
  \bibitem{Chen} J.Z.~Chen and D.~Jia,
                 astro-ph/0512058;
  \bibitem{Sundrum} R.~Sundrum,
                    \prd{59}{1999}{085009}, hep-ph/9805471;
  \bibitem{Weinberg} S.~Weinberg,
                     {\it Gravitation and Cosmology},
                      Wiley \& Sons, 1972;
  \bibitem{HorWitt2} P.~Ho\v{r}ava and E.~Witten,
                     \npb{475}{96}{94}, hep-th/9603142;
  \bibitem{RS2} L.~Randall and R.~Sundrum,
                \prl{83}{1999}{4690}, hep-th/9906064;
  \bibitem{GKL} G.W.~Gibbons, R.~Kallosh and A.D.~Linde,
                \jhep{0101}{2001}{022}, hep-th/0011225;
  \bibitem{Lykken} J.D.~Lykken, R.C.~Myers and J.~Wang,
                   \jhep{0009}{2001}{009}, hep-th/0006191;
  \bibitem{CPV} C.S.~Chan, P.L.~Paul and H.~Verlinde,
                \npb{581}{2000}{156}, hep-th/0003236;
  \bibitem{Linde} A.~Linde,
                  {\it Particle Physics and Inflationary Cosmology},
                  Harwood Academic, 1990;
  \bibitem{KraHet} A.~Krause,
                   \jhep{0005}{2000}{046}, hep-th/9909182;
                   G.~Curio and A.~Krause,
                   \npb{643}{2002}{131}, hep-th/0108220;
                   A.~Krause,
                   \fp{49}{2001}{163};
  \bibitem{AIQU} G.~Aldazabal, L.E.~Ibanez, F.~Quevedo and A.M.~Uranga,
                 \jhep{0008}{2000}{002}, hep-th/0005067;
  \bibitem{AKT}  I.~Antoniadis, E.~Kiritsis and T.N.~Tomaras,
                 \plb{486}{2000}{186}, hep-ph/0004214;
  \bibitem{SVW} S.~Sethi, C.~Vafa and E.~Witten,
                \npb{480}{1996}{213}, hep-th/9606122;
  \bibitem{GVW} S.~Gukov, C.~Vafa and E.~Witten,
                \npb{584}{2000}{69}, hep-th/9906070;
  \bibitem{BC1} K.~Behrndt and M.~Cvetic,
                \plb{475}{2000}{253}, hep-th/9909058;
  \bibitem{BC2} K.~Behrndt and M.~Cvetic,
                \prd{61}{2000}{101901}, hep-th/0001159;
  \bibitem{KL} R.~Kallosh and A.~Linde,
               \jhep{0002}{2000}{005}, hep-th/0001071;
  \bibitem{CLP} M.~Cvetic, H.~Lu and C.N.~Pope,
                {{\sl Class. Quant. Grav.} {\bf 17} (2000) 4867}, hep-th/0001002;
  \bibitem{ACN} R.~Arnowitt, A.~Chamseddine and P.~Nath,
                \plb{156}{1985}{215};
  \bibitem{Mohapatra} R.N.~Mohapatra,
                      {\it Supersymmetric Grand Unification: An
                      Update},
                      hep-ph/9911272;
  \bibitem{KCT2} A.~Krause,
                 hep-th/0204206;
                 {\sl Proceedings of SUSY 2002}, hep-th/0212339;
                 \cqg{20}{2003}{S533};
  \bibitem{KBH} A.~Krause,
                hep-th/0312309; hep-th/0312311;
  \bibitem{GUT} Z.~Berezhiani and G.~Dvali,
                \splir{5}{1989}{55};
                R.~Barbieri,
                \npb{432}{1994}{49};
                Z.~Berezhiani, L.~Randall and C.~Csaki,
                \npb{444}{1995}{61};
                Z.~Berezhiani,
                \plb{355}{1995}{481};
                G.~Dvali and S.~Pokorski,
                \prl{78}{1997}{807}, hep-ph/9610431;
  \bibitem{Mass1} C.~Vafa and E.~Witten,
                  \npb{431}{1994}{3}, hep-th/9408074;
                  R.~Donagi and E.~Witten,
                  \npb{460}{1996}{299}, hep-th/9510101;
  \bibitem{Mass2} J.~Polchinski and M.J.~Strassler, hep-th/0003136

\end{thebibliography}
\end{document}